Hypothesis Article

# The Role of $N_2$ as a Geo-Biosignature for the Detection and Characterization of Earth-like Habitats

Helmut Lammer[1], Laurenz Sproß[1,2], John Lee Grenfell[3], Manuel Scherf[1], Luca Fossati[1], Monika Lendl[1], Patricio E. Cubillos[1]

[1] Austrian Academy of Sciences, Space Research Institute, Graz, Austria
[2] Institute of Physics, University of Graz, Austria
[3] Department of Extrasolar Planets and Atmospheres, German Aerospace Center, Institute of Planetary Research, Berlin, Germany

Running Title: $N_2$-Atmospheres on Earth-like Planets

Corresponding Author:
helmut.lammer@oeaw.ac.at
Austrian Academy of Sciences, Space Research Institute,
Schmiedlstr. 6, 8042 Graz, Austria
helmut.lammer@oeaw.ac.at

## ABSTRACT

Since the Archean, $N_2$ has been a major atmospheric constituent in Earth's atmosphere. Nitrogen is an essential element in the building blocks of life, therefore the geobiological nitrogen cycle is a fundamental factor in the long term evolution of both Earth and Earth-like exoplanets. We discuss the development of the Earth's $N_2$ atmosphere since the planet's formation and its relation with the geobiological cycle. Then we suggest atmospheric evolution scenarios and their possible interaction with life forms: firstly, for a stagnant-lid anoxic world, secondly for a tectonically active anoxic world, and thirdly for an oxidized tectonically active world. Furthermore, we discuss a possible demise of present Earth's biosphere and its effects on the atmosphere. Since life forms are the most efficient means for recycling deposited nitrogen back into the atmosphere nowadays, they sustain its surface partial pressure at high levels. Also, the simultaneous presence of significant $N_2$ and $O_2$ is chemically incompatible in an atmosphere over geological timescales. Thus, we argue that an $N_2$-dominated atmosphere in combination with $O_2$ on Earth-like planets within circumstellar habitable zones can be considered as a geo-biosignature. Terrestrial planets with such atmospheres will have an operating tectonic regime connected with an aerobe biosphere, whereas other scenarios in most cases end up with a $CO_2$-dominated atmosphere. We conclude with implications for the search for life on Earth-like exoplanets inside the habitable zones of M to K-stars.

## 1. INTRODUCTION

"Are we alone in the universe?" The discovery and characterization of exoplanets around Sun-like stars which began in 1995 (Mayor and Queloz, 1995), is gradually bringing us closer to answering this fundamental question of humanity. There are however two key aspects to consider for finding life as we know it: firstly, we need to detect a large sample of Earth-like planets in their host star's habitable zone and, secondly, we need to detect and confirm biosignatures (e.g. in the form of atmospheric gases).

The evolution of an Earth-like planet and its atmosphere is strongly related to various processes, e.g. the planet's formation, its initial volatile and water inventory, the host star's activity controlling the escape of the planetary protoatmosphere, the evolution of the secondary atmosphere, and the planet's impact history (e.g. Halliday, 2003; Lammer *et al.*, 2013a; Mikhail and Sverjensky, 2014; Wordsworth, 2016; Catling and Kasting, 2017; Zerkle and Mikhail, 2017; Lammer and Blanc, 2018; Lammer *et al.*, 2018).

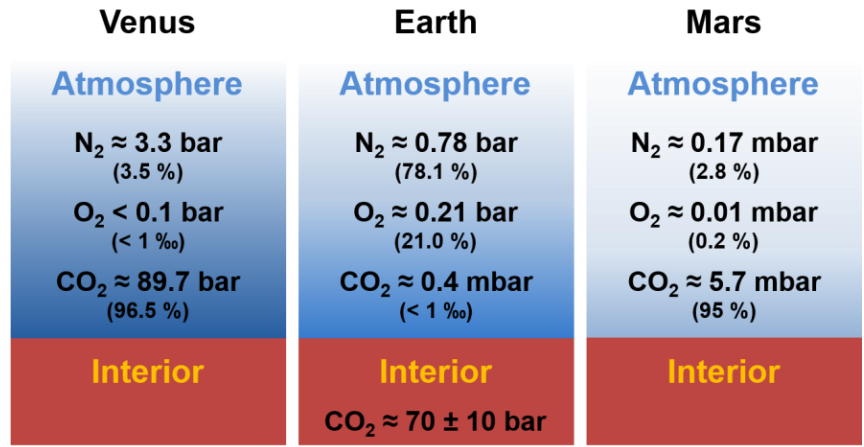

**Fig. 1.** Percentage by volume of total $N_2$, $O_2$ and $CO_2$ contents in the Venusian (Oyama *et al.*, 1980), terrestrial (dry air) (*NASA*, 2017), and Martian (Franz *et al.*, 2017) atmospheres. An atmospheric equivalent of about 60-80 bar $CO_2$ is stored in the Earth's crust in the form of carbonates (e.g. Ronov and Yaroshevskiy, 1967; Holland, 1978; Walker, 1985; Kasting, 1993). The different blue shades represent the different atmospheric densities.

As illustrated in Fig. 1, atmospheric percentages of $CO_2$ and $N_2$ on Venus and Mars would be similar to those of present-day Earth, if Earth had not depleted its atmospheric $CO_2$ through weathering during the Hadean (e.g. Walker, 1985; Kasting, 1993; Sleep and Zahnle, 2001). At some point in history $N_2$ became the dominant constituent in the terrestrial atmosphere. When this happened is a matter of debate, as is the evolution of atmospheric nitrogen. While research on fossilized raindrop imprints suggests that the atmospheric pressure was low in the Archean, probably less than half the present-day value (Som *et al.*, 2012; 2016; Marty *et al.*, 2013; Avice *et al.*, 2018), studies of subduction zones indicate Archean nitrogen partial pressures above the present-day value (Goldblatt *et al.*, 2009; Barry and Hilton, 2016; Johnson and Goldblatt, 2018; Mallik *et al.*, 2018). A better understanding of the evolution of the nitrogen cycle is of crucial importance to address this controversy.

Besides the Earth, our solar system features icy moons or dwarf planets lying far beyond the ice line but also possessing $N_2$-dominated atmospheres. For example, Titan (Strobel and Shemansky, 1982) has a 1.45 bar atmosphere consisting of 98.4 % $N_2$, 1.4 % $CH_4$, and 0.1-0.2 % $H_2$ (Coustenis and Taylor, 2008). However, the origin of such atmospheres is related to early photolysis of accreted and outgassed $NH_3$ from sub-surface $H_2O$-$NH_3$-oceans leading to a very different environment compared to Earth's $N_2$ atmosphere (Coustenis and Taylor, 2008; Mandt *et al.*, 2009; 2014). In this hypothesis paper we focus on classical rocky terrestrial planets that originated in an inner planetary system; frozen worlds like Titan, Triton, and Pluto are not considered. Beside these environments, we are aware of alternatively conditioned habitats (e.g. Jones, 2003; Lammer *et al.*, 2009), however, at the moment we have no evidence for non-Earth biochemistries, thus we focus on Earth-like biospheres.

It is well known that the origin and evolution of life on Earth has a strong influence on Earth's atmospheric composition and climate (Kiehl and Dickinson, 1987; Haqq-Misra *et al.*, 2008; Wolf and Toon, 2013; Kunze *et al.*, 2014; Catling and Kasting, 2017; Charnay *et al.*, 2017). The potentially major role of nitrogen is often overlooked. Recently, Stüeken et al. (2016a) simulated atmospheric-biological interactions over geological times on Earth-like planets and even concluded that $N_2$ and $O_2$ in combination could be a possible signature of an oxygen-producing biosphere. This is also supported by thermodynamic studies (Krissansen-Totton *et al.*, 2016a). One should note that nitrogen is an essential element for all life forms on Earth since it is required, like carbon and phosphorus, for the formation of nucleic acids and proteins.

In order to search for life on extrasolar planets, a set of tell-tale atmospheric signatures (molecular "biosignatures") have been discussed that would allow for the detection and characterization of biospheres (Lovelock, 1975; Segura *et al.*, 2003; Cockell *et al.*, 2009; Kaltenegger *et al.*, 2007; Grenfell *et al.*, 2007a; 2007b; 2010; for a review see Schwieterman *et al.*, 2018). Oxygen is a necessary ingredient for the evolution of complex life forms on habitable planets, as discussed in detail by Catling et al. (2005) and Meadows et al. (2017). $O_2$ has long been recognized as a key-biosignature, detectable by subsequently produced $O_3$ (Owen, 1980; Léger *et al.*, 1993; 2011; Sagan *et al.*, 1993; Des Marais *et al.*, 2002; Airapetian *et al.*, 2017a). However, several more recent theoretical studies have shown that $O_2$ may also build up abiotically in an exoplanet's atmosphere (Domagal-Goldman *et al.*, 2014; Gao *et al.*, 2015; Harman *et al.*, 2015; Luger and Barnes, 2015; Tian *et al.*, 2014; Wordsworth and Pierrehumbert, 2014). Depending on a planet's gravity and the host star's EUV flux evolution, an important pathway for abiotically raising atmospheric oxygen levels consists of $H_2O$ dissociation followed by hydrogen escape (e.g. Zahnle and Kasting, 1986; Lammer *et al.*, 2011; Luger and Barnes, 2015). If the escape of oxygen is considerably less efficient than that of hydrogen, this could lead to the existence of terrestrial habitable-zone planets with high levels of abiotically accumulated $O_2$, as long as processes which potentially deplete atmospheric oxygen (e.g. surface oxidation) are inefficient. Such a scenario can also happen if the liquid ocean of a terrestrial planet is formed after the EUV saturation phase of the host star (Tu *et al.*, 2015), when the decreased stellar EUV flux is insufficient to remove the dense abiotic oxygen atmosphere. Such $O_2$-rich atmospheres will also produce $O_3$ layers that, accompanied by the detection of $H_2O$ and relatively low $CO_2$ values, could result in potential false positives for life. Grenfell et al. (2018) investigated atmospheric $H_2$-$O_2$-combustion as an additional $O_2$-sink and source of water.

Further proposed biosignature molecules include $N_2O$ (Sagan *et al.*, 1993; Segura *et al.*, 2005; Rauer *et al.*, 2011; Rugheimer *et al.*, 2013; 2015; Airapetian *et al.*, 2017a), $CH_4$ (Sagan *et al.*, 1993; Krasnopolsky *et al.*, 2004; Rugheimer *et al.*, 2015; Airapetian *et al.*, 2017a), $CH_3Cl$ (Segura *et al.*, 2005; Rugheimer *et al.*, 2015), $NH_3$ (Seager *et al.*, 2013a; 2013b), sulfur gases and $C_2H_6$ (Pilcher, 2003; Domagal-Goldman *et al.*, 2011), and organic hazes (Arney *et al.*, 2016; 2017). The detection of these species does not necessarily imply that a particular planet is populated by aerobic life forms as we know them. In order to improve our capacity to interpret these signals in their environmental context, it makes sense to investigate the connection between atmospheric oxygen, Earth's main atmospheric species $N_2$, and the evolution of life.

But which scenarios lead to $N_2$-dominated atmospheres on Earth-like planets and which do not? Section 2 investigates the role of atmospheric $N_2$ and its co-existence with life as we know it in more detail, while Section 0 discusses partial pressure on Earth in the past. In Section 4, we discuss four atmospheric development scenarios, based on the above described geobiological interactions, while Section 5 discusses the same for Earth-like planets orbiting M- and K-stars. Possibilities for the detection of $N_2$-dominated Earth-like atmospheres on exoplanets are discussed in Section 6. In Section 7 we conclude under which conditions $N_2$ accompanied by $O_2$ can be a geo-biosignature, here defined as a biosignature that is strongly linked with tectonic activity.

## 2. PROCESSES AFFECTING EARTH'S EARLY ATMOSPHERIC NITROGEN EVOLUTION

### 2.1 Abiotic magmatic surface interactions and steam atmosphere conditions

In the earliest stages of planet formation, protoplanetary cores can accumulate $H_2$-envelopes (Sekiya *et al.*, 1980a; Sekiya *et al.*, 1980b; Sasaki and Nakazawa, 1988; Lammer and Blanc, 2018). Due to the captured nebula gas, serpentinization, and accreting chondritic material, crust and atmosphere were strongly reducing and later oxidized during the planet's life (Schaefer and Fegley, 2010). Hydrogen is partly lost during the short very efficient thermal escape phase ("boil-off phase") and can be completely removed due to EUV-driven hydrodynamic escape (e.g. Gillmann *et al.*, 2009; Lammer *et al.*, 2014; 2018; Johnstone *et al.*, 2015; Fossati *et al.*, 2017; Odert *et al.*, 2018). After the escape of the nebular gas, the deep magma ocean on the planet's surface solidifies and thereby outgasses $H_2O$ and $CO_2$ catastrophically, so that a dense steam atmosphere evolves (Sleep, 2010). Depending on the magma ocean's lifetime as well as the cooling time of this atmosphere, water is either partly lost to space or later condenses to form an ocean (e.g. Elkins-Tanton, 2008; 2012; Hamano *et al.*, 2013; Lebrun *et al.*, 2013; Massol *et al.*, 2016; Salvador *et al.*, 2017).

Due to the catastrophic outgassing during to the magma ocean phase, the nitrogen partial pressure could have reached a few 100s of mbar (Holland, 1984; Turner *et al.*, 1990). Assuming 70 bar of outgassed $CO_2$, plausible for the Earth case, and Earth's C/N ratio of present mid-ocean ridge outgassing obtained by various studies (Zhang and Zindler, 1993; Marty, 1995; Marty and Tolstikhin, 1998; Sano *et al.*, 2001; Coltice *et al.*, 2004; Cartigny *et al.*, 2008; Marty *et al.*, 2013) the total outgassed $N_2$ is in the range of 24-203 mbar (for 500 mbar $N_2$ at least ~170 bar of $CO_2$ would have to be outgassed). After the outgassing phase, under the still extremely hot surface conditions (>1000 K) in combination with a reducing steam environment, efficient atmospheric $NH_3$ production occurs (Schaefer and Fegley, 2010; Wordsworth, 2016) at a rate which outpaces the dissociation by FUV and EUV radiation (e.g. Holland, 1962; Kuhn and Atreya, 1979; Kasting, 1982; 1993; Zahnle *et al.*, 2013). There are indications by experimental studies that massive direct dissolution into the mantle is then possible in such reduced environments (Solomatov, 2000; Libourel *et al.*, 2003; Kadik *et al.*, 2011). Thus, one can expect that the majority of atmospheric nitrogen is quickly sequestered into the hot surface environment, as illustrated in Fig. 2a (Wordsworth, 2016).

After the magmatic mantle solidified, the steam in the atmosphere eventually condenses to produce produces a warm liquid $H_2O$ ocean while the atmosphere contains several tens of bar $CO_2$ (e.g. Ronov and Yaroshevskiy, 1967; Holland, 1978; Kasting, 1993; Zahnle, 2006; Lammer et al., 2018). Then, as illustrated in Fig. 2b, remaining atmospheric nitrogen is still affected by diverse abiotic fixation processes (see also Section 0; for comparison, Earth's present lightning fixes 10 mbar N in ~10 Myr).

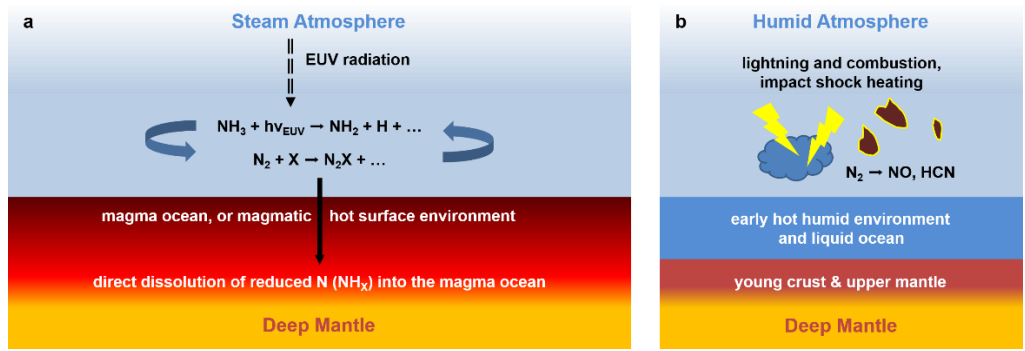

**Fig. 2.** a) Abiotic atmospheric-surface weathering processes capable of transferring atmospheric nitrogen into early Earth's surface and mantle via direct dissolution of reduced nitrogen into the early magma ocean(s). This efficient abiotic mechanism most likely operated on early Venus and Earth during and just after accretion. 'X' represents a reducing species, such as H (adapted from Wordsworth, 2016). b) Additionally to the magmatic weathering process, atmospheric $N_2$ will also undergo fixation via weathering caused by lightning, shock heating via impactors, and energetic particles in the early steam atmosphere.

## 2.2 Nitrogen speciation in Earth's upper mantle

Diverse studies of Earth's $N_2$ atmospheric and interior inventories indicate that outgassing of $N_2$ in the first Gyr was strongly connected to the planet's thermodynamic evolution (e.g. Busigny and Bebout, 2013; Mikhail and Sverjensky, 2014; Wordsworth, 2016; Zerkle and Mikhail, 2017) and oxidation stages of crust and upper mantle (Kasting, 1993; Delano, 2001; Catling and Claire, 2005; Kelley and Cottrell, 2009; Trail *et al.*, 2011; Catling and Kasting, 2017;

Zerkle *et al.*, 2017). Mikhail and Sverjensky (2014) found out that the speciation of $N_2$ in high-pressure, supercritical aqueous fluids in Earth's mantle wedge are the most likely origin of Earth's $N_2$ atmosphere.

Molecular nitrogen is highly incompatible in silicate minerals (Li *et al.*, 2013), while ammonic nitrogen can be moderately compatible in silicates like phlogopite and clinopyroxene (e.g. Watenphul *et al.*, 2010; Li *et al.*, 2013; Mikhail and Sverjensky, 2014; Zerkle and Mikhail, 2017). This is in agreement with experimental data, which indicate that under very oxidizing environmental conditions nitrogen in supercritical fluids remains as dinitrogen ($N_2$) whereas it exists as $NH_3$ under reducing conditions (Canfield *et al.*, 2010). Thermodynamic studies of Mikhail and Sverjensky (2014) indicate that Earth's upper mantle nitrogen inventory is usually present in the form of ammonium ($NH_4^+$) in aqueous fluids and upper mantle minerals. Since Earth developed tectonic activity, subduction of oceanic lithosphere carries oxidized surface rocks and large amounts of water into Earth's upper mantle (e.g. McCammon, 2005; Hirschmann, 2009; Lammer *et al.*, 2018). This can locally change the redox state to favor $N_2$, which is easily outgassed. Estimates for the start of this process range from 4.35 Gyr (Trail *et al.*, 2011) ago to 3.8 Gyr ago (Delano, 2001).

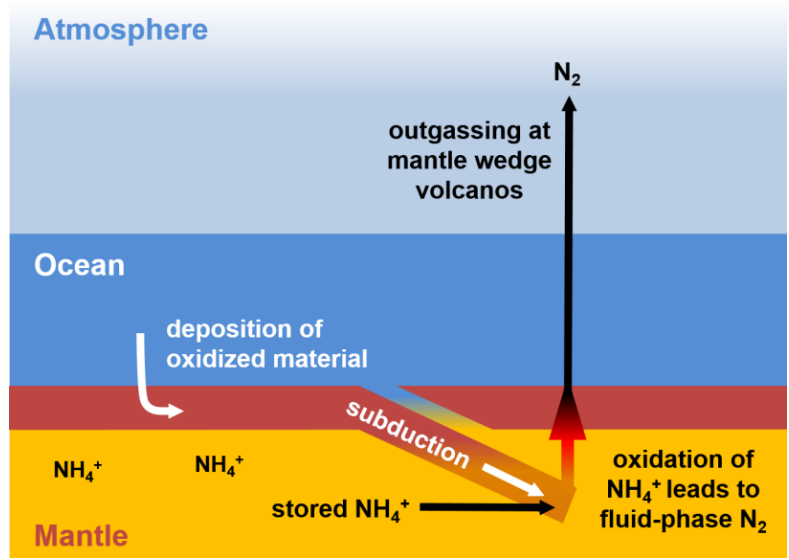

**Fig. 3.** Schematic vertical cross-section through a subduction zone, displaying the geochemical cycle of fluids and outgassing related to mantle wedge areas. Oxidized material is transported via subduction zones into the crust and upper mantle where it reacts with $NH_4^+$ and is hence efficiently outgassed from mantle wedge volcanos in the form of $N_2$.

## 2.3 Development of the geobiological nitrogen cycle

The geobiological nitrogen cycle shows different features before and after oxygen rose in the atmosphere. The major processes affecting nitrogen in both cases are illustrated in Fig. 4. Molecular nitrogen, which was outgassed into the Earth's early atmosphere, is chemically inert. Any nitrogen fixation process which could convert nitrogen into more chemically reactive compounds, requires high energy.

Abiotic fixation in the early Archean included lightning, high energy particle interaction, atmospheric shock heating by frequent meteorite impacts, a higher solar ultraviolet radiation, and coronal mass ejections related to super flares (e.g. Airapetian *et al.*, 2016). Generally, also throughout later periods until today, non-biological pathways occur via high-temperature reducing or oxidation reactions of $N_2$ to $NH_x/HCN/NO_x$ (Navarro-González *et al.*, 2001; Martin *et al.*, 2007; Parkos *et al.*, 2016), depending on the environment's redox state. These occur during combustion or lightning in the troposphere, followed by conversion into water-soluble molecules (e.g. $HNO_3$) within the atmosphere, which are quickly scavenged by rain.

There is also a biotic fixation pathway since some bacteria are able to reduce atmospheric $N_2$ to $NH_3$ (biological nitrogen fixation, often abbreviated as BNF). Since nitrogen is an ingredient for the building blocks of life, this reduced nitrogen is often assimilated as organic nitrogen ($N_{org}$) by microorganisms and, in more recent times, by plants. Both may also be eaten by other lifeforms who either excrete the nitrogen or release it after death. Afterwards, this $N_{org}$ is again consumed by bacteria and mineralized into ammonium $NH_4^+$ that can be assimilated again by other organisms (e.g. Boyd and Philippot, 1998; Boyd, 2001; Holloway and Dahlgren, 2002; Mikhail and Sverjensky, 2014; Wordsworth, 2016; Zerkle and Mikhail, 2017).

After oxygen enriched the atmosphere during the Great Oxidation Event (GOE), bacteria have also used $NH_4^+$ as a source of energy on a large scale by oxidizing it to nitrite $NO_2^-$ and to nitrate $NO_3^-$, a process called nitrification (Fig. 4b; Jacob, 1999; Zerkle and Mikhail, 2017; Zerkle *et al.*, 2017). Some bioavailable nitrogen in the Earth's ocean is returned to the atmosphere as $N_2$ via denitrification, which means the reduction of nitrate $NO_3^-$ to $N_2$. This microbially facilitated process is performed by heterotrophic bacteria such as Paracoccus denitrificans.

Beside these processes, under anaerobic conditions where molecular oxygen is depleted, bacteria can use nitrates as an alternate oxidant to convert organic carbon into $CO_2$ while releasing $N_2$. Another biological process that releases $N_2$ into the atmosphere is called anaerobic ammonium oxidation (anammox), which is the oxidation of $NH_4^+$ with nitrite ($NO_2^-$) that are converted directly into diatomic nitrogen and water. Globally, this process may be responsible for 30 - 50 % of the $N_2$ gas produced in the oceans, which is then released into the atmosphere (Devol, 2003). It is not clear as to when this process started to play a role in the nitrogen cycle and it might have been negligible up to the GOE (Som *et al.*, 2016). The requirements for anammox however are fulfilled as early as the late Archean (Stüeken *et al.*, 2016b).

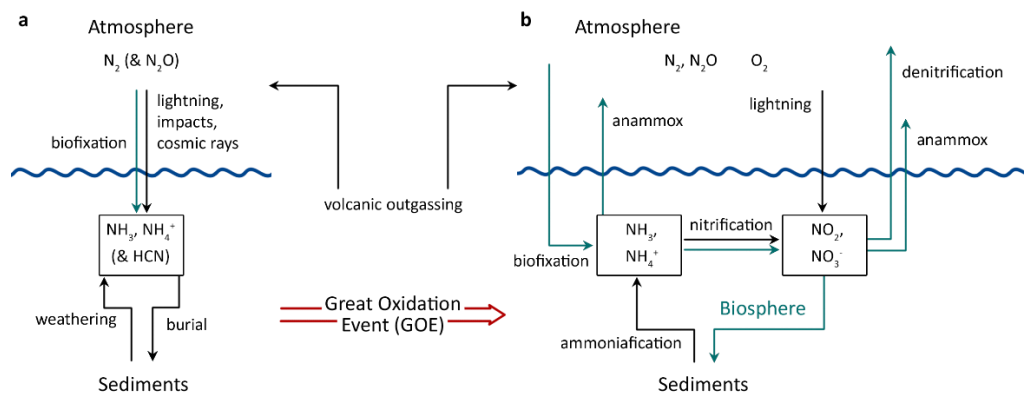

**Fig. 4.** Illustrations of the major nitrogen processes during the Archean and Proterozoic i.e. before and after the Great Oxidation Event (GOE) took place, approximately 2.3 Gyr ago. Black and green arrows indicate abiotic and biotic processes, respectively. a) Biogeochemical nitrogen cycle in the Archean before the GOE when anaerobic ammonium oxidation (anammox) and denitrification did not take place. b) As for a) but after the GOE, a change occurred when oxidation of ammonia to nitrate ($NO_3^-$) and nitrite ($NO_2^-$), so-called "nitrification" set in. These reactions have resulted in the necessary substrate for the reduction of nitrate to atmospheric $N_2$ (denitrification) and anammox to atmospheric $N_2$ via nitrite.

## 2.4 Biogenic (and anthropogenic) influences on present Earth's nitrogen cycle

Although $N_2$ is photochemically inert, nitrogen in the Earth's system was (and still is) efficiently cycled by life forms (e.g., phytoplankton, cyanobacteria, etc.) at a rate of about $2 \times 10^{14}$ g N/yr (200 Tg N/yr) (e.g. Schlesinger, 1997; Jacob, 1999;

Galloway, 2003; Cartigny and Marty, 2013), as discussed in Section 2.3 and shown in Fig. 4b. A summary of present total exchange rates of atmospheric nitrogen is provided in Tab. 1. Tables in the appendix list exchange rates as estimated by various studies.

Today's net nitrogen flux is not undisputed; it is not even clear if there is a net out- or ingassing of nitrogen on Earth (Zerkle and Mikhail, 2017). The massive anthropogenic influence on this system further leads to uncertainties on any conclusion that could be drawn. Human influence destabilizes the well-balanced nitrogen cycle, such that soils and oceans are overloaded with atmospheric nitrogen. This might also be a reason for the rise in $N_2O$ simultaneous to the anthropogenic nitrification over the last decades, which indicates a rise in denitrification. One should note that Earth is a highly dynamic planet with active plate tectonics, and varying volcanic activity. The exchange of nitrogen between the atmosphere and the planet's interior is controlled by subduction and volcanism (e.g. Sano *et al.*, 2001; Fischer, 2008). Therefore, the efficiency of nitrogen deposition and outgassing processes which control the atmospheric $N_2$ partial surface pressure is dependent on the Earth's inner dynamics and cannot be used linearly for long-time estimates for past or future conditions (making also the studies by Barry and Hilton (2016) and Mallik et al. (2018) debatable).

**Tab. 1.** Mass budget and source sink inventory estimates in the present Earth's upper nitrogen cycle, consisting of atmosphere, land biota, soil and ocean biota from the atmosphere into the surface/interior ("-") and from the surface into the atmosphere ("+"). The lower part of the nitrogen cycle including volcanic processes contains rates which are estimated to be at least one order of magnitude smaller than the smallest rates in this table. All values are given in Tg.

| Atmosphere $3.95 \times 10^9$ [a] | Soil [b] $1.0 \times 10^9$ [a] | Land Biota $1.0 \times 10^4$ [a] | Ocean $2.06 \times 10^7$ [a] | Ocean biota $5.0 \times 10^2$ [a] |
|---|---|---|---|---|
| Biofixation | | $-1.2 \times 10^2$ yr$^{-1}$ [c] | | $-1.4 \times 10^2$ yr$^{-1}$ [c] |
| Rain | $-7.0 \times 10^1$ yr$^{-1}$ [c] | | $-3.0 \times 10^1$ yr$^{-1}$ [c] | |
| Denitrification | | $+1.1 \times 10^2$ yr$^{-1}$ [c] | | $+1.9 \times 10^2$ yr$^{-1}$ [c] |
| Biomass burning | | $+5.0$ yr$^{-1}$ [c] | | |
| Surface release | $+7.8 \times 10^1$ yr$^{-1}$ [c] | | $+1.5 \times 10^1$ yr$^{-1}$ [c] | |
| Industry (fertilizers) | $-1.2 \times 10^2$ yr$^{-1}$ [c] | | | |

[a] (Galloway, 2003)

[b] The amount of organic nitrogen in soils is estimated to be $2 \times 10^5$.

[c] (Fowler *et al.*, 2013)

For an overview of different rate estimates see tables in the appendix.

It is important to note that the present day volcanic outgassing rates of (1.5 ± 1.0) Tg N/yr (Tab. 5) cannot balance the atmospheric $N_2$ sinks by fixation of (430 ± 60) Tg N/yr (Tab. 1,

Tab. **2**). Also if one subtracts the anthropogenic influence, the remaining removal flux is still far too high to be balanced by present volcanic outgassing. Without biological denitrification and anammox that are responsible for a strong return flux, Earth's present-day atmospheric nitrogen of approximately $4 \times 10^9$ Tg would be entirely sequestered within less than 100 Myr (e.g. Cartigny and Marty, 2013; Lammer *et al.*, 2018). Thus, one can conclude that the present day partial surface pressure of about 0.78 bar is mainly maintained by bacteria, which under anaerobic conditions, return $N_2$ from the surface environment to the atmosphere (e.g. Jacob, 1999; Cartigny and Marty, 2013; Wordsworth, 2016; Zerkle *et al.*, 2017).

## 3. THE LIKELY EVOLUTION OF THE NITROGEN PARTIAL PRESSURE ON EARTH

Taking the considerations made in Section 2.1 into account, for Earth one can assume some tens of mbar nitrogen to be outgassed from the final magma ocean in the Hadean. In the hot $CO_2/H_2O$ atmosphere $N_2$ can be efficiently converted into $NH_3$ which subsequently sequesters the majority of nitrogen back into the hot surface environment (Wordsworth, 2016). Abiotic fixation processes such as lightning (e.g. Chameides and Walker, 1981; Navarro-González *et al.*, 2001) that was most likely efficient in a dense humid atmosphere, EUV-related photochemistry (Zahnle, 1986; Tian *et al.*, 2011; Airapetian *et al.*, 2016), frequent meteoritic impacts (Fegley *et al.*, 1986; Chyba and Sagan, 1992; Parkos *et al.*, 2018) and cosmic rays (Navarro-González *et al.*, 1998; Grenfell *et al.*, 2012; Cooray, 2015; Tabataba-Vakili *et al.*, 2016) provided additional energy to fix nitrogen from the atmosphere (see Fig. 2b), even during the post-magmatic surface period. In a scenario where, despite efficient fixation, a relatively high nitrogen partial pressure remains, nitrogen would suffer strong atmospheric escape in cases where it is more abundant than $CO_2$ that cools the thermosphere and hinders escape (see Fig. 9; Tian *et al.*, 2008a; Lichtenegger *et al.*, 2010; Lammer *et al.*, 2011; 2018). The absence of a footprint of such an escape in the atmospheric $^{14}N/^{15}N$ isotope ratio on Earth indicates only percentage levels of $N_2$ compared to $CO_2$ to have been present in the late Hadean (Lichtenegger *et al.*, 2010; Cartigny and Marty, 2013; Avice *et al.*, 2018; Lammer *et al.*, 2018).

All these arguments make nitrogen partial pressures as high as today (or even higher), as assumed in some studies (Goldblatt *et al.*, 2009; Johnson and Goldblatt, 2018; Barry and Hilton, 2016; Mallik *et al.*, 2018), unlikely for this period (Lammer *et al.*, 2008; 2011; 2013a; 2018; Tian *et al.*, 2008a; 2008b; Lichtenegger *et al.*, 2010; Scherf *et al.*, 2018). Moreover, higher outgassing fluxes (e.g. Fischer, 2008; our Tab. 5) that are comparable to subduction fluxes counteract the arguments for a high initial $N_2$ partial pressure presented by Barry and Hilton (2016) and Mallik et al. (2018). Thus, a buildup of $N_2$ in the Archean is plausible.

Also during the Archean, one can assume a high number of charge carriers to be present over the wide water ocean surface. Under these conditions lightning fixation can be efficient (e.g. Rakov and Uman, 2003; 2004; Cooray, 2015). On present Earth, this fixation is estimated to be $(4 \pm 1)$ Tg N/yr (

Tab. **2**), depleting 100 mbar $N_2$ in 100 Myr. One should be aware that a linear dependency of $NO_x$ production rate to air pressure, as frequently used in models, is not accurate (Navarro-González *et al.*, 2001), therefore this process might often be underestimated for early Earth. Moreover, if one assumes higher volcanic activity followed by intense discharges in the outgassed $H_2O$-$CO_2$-rich mixture of gases, an additional fixation pathway opens up (Navarro-González *et al.*, 1998). The other above mentioned fixation processes are still present in the Archean and estimated to fix 1-10 Tg N/yr (Navarro-González *et al.*, 1998), whereby EUV-driven photochemistry and the impactor flux decrease in efficiency over time. All these processes can also be responsible for substantial amounts of HCN molecules in a reducing atmosphere, quickly deposited by rain (e.g. Zahnle, 1986; Parkos *et al.*, 2016; Martin *et al.*, 2007). Assuming today's nitrogen outgassing, the buildup of a significant partial surface pressure within this period is not possible - even assuming a factor of 10 higher volcanic activities (Sano *et al.*, 2001; Hilton *et al.*, 2002, see tables in the appendix). Low pressure scenarios are also supported by studies of Marty et al. (2013) and Som et al. (2016), which indicate that the Archean atmosphere had a total surface pressure of 0.23 - 0.5 bar or even lower.

After the lithospheric oxidation closed up on the GOE-level, the nitrogen partial pressure rose dramatically because heterotrophic microorganisms capable of denitrification (or anammox) began to release $N_2$ and therefore to counteract fixation (see Section 2.3). Since this biological recycling involves oxygen, the buildup of a dense $N_2$ atmosphere on early Earth can be correlated with the rise of atmospheric $O_2$ shortly before and during the GOE (e.g. Catling *et al.*, 2005; Lyons *et al.*, 2014; Catling and Kasting, 2017; Lammer *et al.*, 2018).

In summary it can be said that Earth's atmosphere during the late Hadean/early Archean was $CO_2$-dominated but with $CO_2$ decreasing over time (Hessler *et al.*, 2004; Kanzaki and Murakami, 2015). As soon as the reduced nitrogen in the upper mantle could be oxidized through subduction (Catling *et al.*, 2001; Kump *et al.*, 2001; Lyons *et al.*, 2014; Aulbach and Stagno, 2016), $N_2$ was outgassed via volcanos at C/N ratios comparable to that of today. After the rise of life, greenhouse gases such as $CH_4$, $N_2O$ (e.g. Catling *et al.*, 2001; Airapetian *et al.*, 2016; Catling and Kasting, 2017; Lammer *et al.*, 2018) and a still substantial amount of $CO_2$ (Kanzaki and Murakami, 2015) kept the surface environment above freezing, which is also indicated by recent 3D global circulation models (Wolf and Toon, 2013; Charnay *et al.*, 2017). Thus, high atmospheric pressures in order to warm the surface by pressure broadening (e.g. Goldblatt *et al.*, 2009; Johnson and Goldblatt, 2018) are not necessary. Finally, at the GOE, $O_2$ built up catastrophically in the atmosphere and the $N_2$ geobiological cycle changed to its modern form (e.g. Zerkle and Mikhail, 2017; Zerkle *et al.*, 2017), so that the surface pressure rose to the present level. The high nitrogen partial pressure on Earth is then directly linked to its biosphere but necessarily combined with oxygen as another bulk gas.

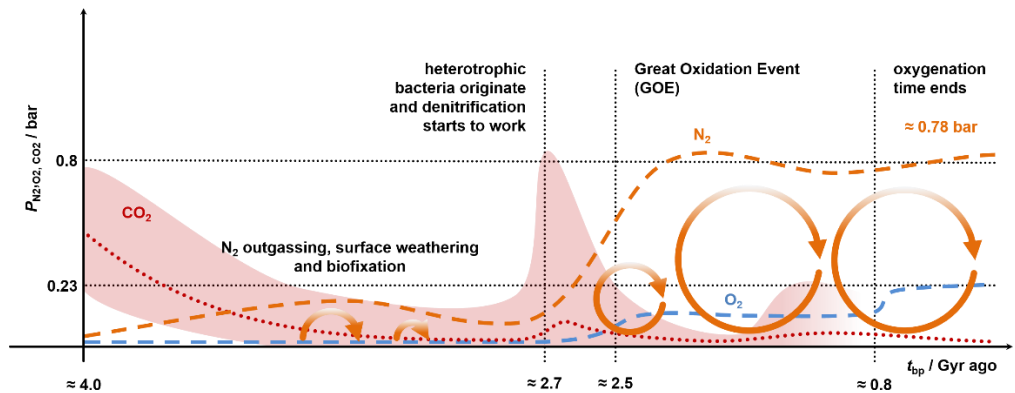

**Fig. 5.** Illustration of the $CO_2$, $N_2$, $O_2$ surface partial pressure evolution on early Earth since about 4 Gyr ago. The ranges for $CO_2$ in light red follow roughly the measurements for 2.77 - 1.85 Gyr by Kanzaki and Murakami (2015). Before about 4 Gyr ago, nitrogen was mainly sedimented in the oceans and stored as $NH_4^+$. After the crust and upper mantle environment became oxidized, nitrogen in the form of $N_2$ was efficiently released into Earth's atmosphere via mantle wedge volcanism above subduction zones (e.g. Mikhail and Sverjensky, 2014; Zerkle and Mikhail, 2017). In the later Archean biological fixation may lowered the partial pressure. $N_2$ then rose to the present values, when heterotrophic microorganisms responsible for denitrification led to the modern geobiological nitrogen cycle. This pressure jump was related to the Great Oxidation Event (GOE) when oxygen was effectively released into the atmosphere (e.g. Catling *et al.*, 2005; Lyons *et al.*, 2014; Catling and Kasting, 2017; Lammer *et al.*, 2018). The semicircles and the full circles illustrate the nitrogen cycle in the early form (drawn in Fig. 4a) respectively in the completed form (Fig. 4b).

# 4. HYPOTHETICAL SCENARIOS FOR THE EVOLUTION OF N$_2$ ATMOSPHERES

The following scenarios consider hypothetical terrestrial planets which have accreted a mass and size similar to those of the Earth, so that one can expect that the protoplanetary core did not accumulate a huge amount of nebular gas which would not be lost during the planet's lifetime (e.g. Lammer *et al.*, 2014; 2018; Johnstone *et al.*, 2015; Owen and Mohanty, 2016; Fossati *et al.*, 2017; Lehmer and Catling, 2017; Lammer and Blanc, 2018). We further assume that the planets orbit around a Sun-like star inside the habitable zone.

After the majority of a surrounding H$_2$-envelope is lost, one can expect that volatiles (H$_2$O, CO$_2$, CO, NH$_3$, HCN, etc.), which have been delivered in the early phase of the accretion via chondrites from the outer planetary system, can be outgassed from a magma ocean formed at the protoplanetary surface. When the final magma ocean solidifies a dense steam atmosphere builds up until the water vapor condenses after 1 - 2 Myr and liquid oceans form (Elkins-Tanton, 2012; Hamano *et al.*, 2013; Lebrun *et al.*, 2013; Massol *et al.*, 2016; Salvador *et al.*, 2017).

A large fraction of the accreted water on a planet inside the habitable zone will eventually exist in liquid form on the planet's surface and in its interior which then features a hydrous mantle transition zone (Pearson *et al.*, 2014; Schmandt *et al.*, 2014; Plümper *et al.*, 2017). In addition to a suitable amount of short-lived radioactive isotopes, water influences the thermal evolution and rock mechanics in the planet's mantle during the later evolutionary stages, including the possibility as to whether plate-tectonics will start to operate or not (e.g. Hopkins and Manning, 2008; Hopkins *et al.*, 2010; Shirey *et al.*, 2008; Korenaga, 2013).

Based on the above evolution scenarios, we consider the following cases and their impact on the evolution of $N_2$ atmospheres:

4.1 Stagnant-lid regime world: Neither plate tectonics nor life evolve, although the planet has a liquid water ocean on its surface.

4.2 Anoxic tectonic world: Plate tectonics evolve and a liquid water ocean is situated on the planet's surface, but no life or only anoxic life forms originate.

4.3 Oxic tectonic (Earth-analogue) world: Origin and evolution scenario as expected to have occurred on the Earth up to the present day.

4.4 Entirely extinct world: All conditions similar to present Earth, but all life forms suddenly become extinct.

## 4.1 Stagnant-lid regime world

In the first scenario, we investigate how the Earth's atmosphere may have evolved if the initial conditions in the early mantle did not favor plate tectonics. Plate tectonics is considered crucial for maintaining the activity of the carbon-silicate cycle over geological timescales hence stabilizing Earth's climate. As discussed above (Section 2.2), the absence of subduction on a terrestrial planet has a profound influence on the chemistry in the lithosphere, which affects outgassing and hence the atmosphere. Fig. 6 illustrates the evolution of such a planet's atmosphere.

After the evaporation of a possible thin nebular-based $H_2$ envelope, a steam atmosphere formed by catastrophic outgassing during magma ocean solidification would have contained mainly $H_2O$, $CO_2$ and, to a lesser extent, nitrogen (e.g. Elkins-Tanton, 2008; 2012). One can expect that atmospheric nitrogen was efficiently weathered in this environment via fixation by lightning, meteoritic impactors and cosmic rays, as described in Section 2.1.

Contrary to a Venus-like case at 0.7 AU, where $H_2O$ likely remained in vapor form, at 1 AU water vapor likely condensed to form a liquid ocean within ~2 Myr (Hamano *et al.*, 2013; Lebrun *et al.*, 2013; Massol *et al.*, 2016; Salvador *et al.*, 2017). Depending on pH level, surface temperature, pressure and alkalinity, such oceans can dissolve atmospheric $CO_2$ leading to possible seafloor weathering (Walker *et al.*, 1981; Pierrehumbert, 2010; Kitzmann *et al.*, 2015; Krissansen-Totton and Catling, 2017; Krissansen-Totton *et al.*, 2018a; Coogan and Gillis, 2018). In the case of a 1 bar $CO_2$ atmosphere, some 30-300 mbar of $CO_2$ remain in the atmosphere (Walker, 1985; Jacob, 1999). Furthermore, $CO_2$ can additionally be weathered via processes such as sequestration of carbon dioxides by carbonate minerals (e.g. Alt and Teagle, 1999; van Berk *et al.*, 2012; Tosi *et al.*, 2017). As illustrated in Fig. 6, atmospheric $CO_2$ therefore decreases during the first few hundred Myr.

Since $CO_2$ is a greenhouse gas and thermospheric IR-cooler it hinders strong thermal atmospheric escape of nitrogen during the EUV active phase of the young Sun/star. If the $CO_2$ pressure drops to values below those of $N_2$, escape to space can emerge as described in Section 0.

Tosi et al. (2017) modeled the outgassing of $CO_2$ and $H_2O$ of a "stagnant-lid Earth", but did not include a magma ocean or the catastrophically-outgassed steam atmosphere (Elkins-Tanton, 2008; 2012; Hamano *et al.*, 2013; Lebrun *et al.*, 2013; Massol *et al.*, 2016). In their model scenarios, secondary outgassed $CO_2$ builds up to surface pressures of ~1.5 bar or less (through weathering) for reducing conditions in the upper mantle (Tosi *et al.*, 2017). For oxidizing conditions, they obtained Venus-like $CO_2$ atmospheres with surface pressures between 100 and 200 bar.

Depending on the efficiency of abiotic $CO_2$ atmosphere-ocean/surface weathering processes, it is possible that the oxidation state on stagnant-lid Earths remains very low due to the missing subduction and related recycling of water and transport of oxidized material into the lower crust and upper mantle (e.g. McCammon, 2005; Tosi *et al.*, 2017). In the case of a Venus-like or stagnant-lid Earth-like planet at closer orbital separations, $H_2O$ may never condense, i.e. no liquid oceans on the planet's surface. If the young host star's EUV-flux is not too high, some residual $O_2$ may remain in the atmosphere as a product of $H_2O$ dissociation, while the hydrogen atoms escape hydrodynamically (Zahnle and Kasting, 1986). The remaining oxygen and atmospheric nitrogen could have been incorporated into the planet's hot magmatic crust, where it oxidized the upper mantle (Gillmann *et al.*, 2009; Hamano *et al.*, 2013; Kurosawa, 2015; Lichtenegger *et al.*, 2016; Wordsworth, 2016; Lammer *et al.*, 2018). Under such conditions, the highly oxidized surface material can be mixed with reduced, nitrogen-rich material. This implies a change in carbon from a reduced form into a more volatile form, which is more easily outgassed. The oxidation at the surface depleting residual oxygen from hydrodynamic escape can potentially explain the 3.4 times greater atmospheric $N_2$ inventory on Venus compared to the Earth (Wordsworth, 2016).

In the absence of this residual oxygen from the escaping $H_2O$-related hydrogen, as illustrated in Fig. 6, no fast and efficient oxidation stage evolves. Due to this – as well as the above mentioned absence of subduction zones and efficient re-introduction of oxidized material into the upper mantle – the secondary outgassed $CO_2$ abundance most likely remains much smaller (i.e., $\leq 1$ bar) on an Earth-like compared to a Venus-like planet (Tosi *et al.*, 2017). Their results further suggested that at 1 AU surface temperatures generally allow the presence of liquid water over almost the entire planetary lifetime. The outer and inner edge of the habitable zone in such stagnant-lid regime worlds is mainly influenced by the amount of outgassed $CO_2$ and other possible greenhouse gases. As mentioned above, these initially $CO_2$-dominated atmospheres can partly dissolve into an ocean.

In conclusion, without massive cycling and a return flux by biological activity atmospheric nitrogen would be removed to the surface and converted to $NH_4^+$ within a few tens of Myr (Galloway, 2003; Cartigny and Marty, 2013; Lammer *et al.*, 2018). Since any active tectonics is assumed to be absent in this scenario, oxidized material does not easily come into contact with the $NH_4^+$ in the planet's interior. Therefore, similar to $CO_2$, no efficient outgassing of $N_2$ occurs so that the planetary atmosphere evolves to be thin with partial pressures in the range of a few mbar to tens of mbar, dominated by $CO_2$ and $N_2$.

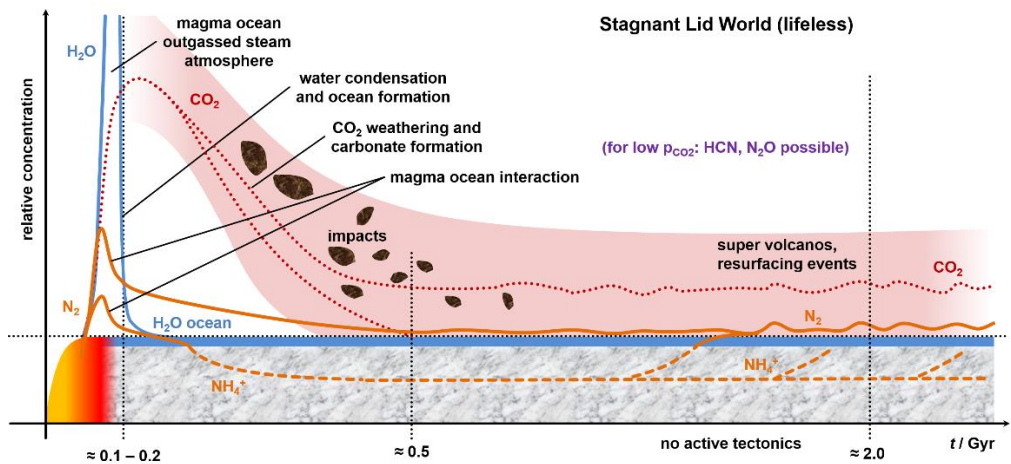

**Fig. 6.** Illustration of the atmospheric evolution of $N_2$ (orange), $CO_2$ (red), and $H_2O$ (blue) for a stagnant-lid regime world. To show the diversity of the pathways for both $N_2$ and $CO_2$ scenarios (lines) are respectively drawn; for $CO_2$ a range is provided (red shaded area). After surface temperatures had dropped, $H_2O$ condensed and formed a liquid ocean. $N_2$ and $CO_2$ are weathered out, while a fraction of the $CO_2$ remains in the atmosphere. Oxidation of the crust and upper mantle is inefficient in such a scenario and oxidized material does not easily get in contact with the $NH_4^+$ in the planet's interior. No efficient outgassing of $CO_2$ and $N_2$ occurs and the planet most likely evolves a thin Martian-like atmosphere whose concentration and related surface pressure may fluctuate depending on the efficiency of $N_2$ and $CO_2$ atmosphere-surface weathering, volcanic activity, availability of additional greenhouse gases (e.g., $N_2O$, $H_2O$, $CH_4$) and possible resurfacing events.

## 4.2 Anoxic tectonic world

In the second scenario, we assume an Earth-like planet, which has developed active tectonics, but either no life, or only anoxic lifeforms originated. Fig. 7 illustrates the planet's atmospheric evolution under such an assumption.

Similar to the stagnant-lid case, a steam atmosphere related to the magma ocean solidification process that catastrophically outgasses (mainly) $H_2O$, $CO_2$ and (to a lesser extent) $N_2$. Then, water vapor condenses after < 2 Myr (at 1 AU) followed by atmosphere-surface interaction processes which weather most of the $CO_2$ and the $N_2$ from the humid atmosphere.

Due to active tectonics, the planet's lithosphere will be gradually oxidized in this scenario, because the oxidized surface material can efficiently enter the crust and upper mantle. Nitrogen can then significantly be released via arc-volcanos in the form of $N_2$ (e.g. Mikhail and Sverjensky, 2014). This can counteract abiotic fixation such as by lightning and impacts (described in Section 2.3) and lead to the buildup of nitrogen as a bulk gas, even with pressures higher than those of $CO_2$. According to Chamaides and Walker (1981) a key parameter that determines the fixation for atmospheric nitrogen is then the ratio of C to O atoms in the atmosphere. Atmospheres with the C/O abundance ratios > 1 have large HCN fixation rates compared to NO yields and vice versa.

Anoxic lifeforms capable of fixing nitrogen can result in its significant atmospheric depletion. According to estimates given in Jacob (1999) this depletion has time scales up to 15 Myr. If only anoxic life originates then nitrogen fixation by lifeforms, lightning and cosmic rays will be dominant and, because of the absence of denitrification, a nitrogen partial pressure similar to that of today is not likely to build up over geological time spans.

Stüeken et al. (2016a) simulated similar cases, but with the assumption of a) today's $N_2$ partial pressure at the very beginning of the planet's origin and b) its total recovery after being depleted. Both assumptions may not be realistic scenarios; a) discussed in Section 2.2 and Fig. 2, b) discussed above, in Section 2.4, and Tab. 1. Also a model by Laneuville et al. (2018) resulted in high atmospheric nitrogen partial pressures (0.5-8 PAL) for a completely lifeless world. One should note that they assumed an active carbon-silicate cycle like on modern Earth and rather low abiotic fixation rates (see lightning rate discussion in Section 0). Nevertheless, their study suggests possible pathways for an abiotic $N_2$-atmosphere with plate tectonics but without $O_2$. In case of anoxic life, as illustrated in Fig. 7, biogenic fixation lowers the nitrogen partial pressure.

According to Kasting et al. (1993), nitrogen reducing at mid-ocean ridges could release $N_2$ back into the atmosphere. For such a process one needs necessarily modern-style plate tectonics. However, this should be modeled in detail to investigate, if it is efficient enough to keep $N_2$ as a major atmospheric constituent. Nevertheless, the $O_2$ surface partial pressure would not rise in such a scenario.

Finally, under the conditions assumed here, an atmosphere will most likely evolve to a thin $CO_2$-dominated atmosphere similar to the stagnant-lid scenario, but with likely less fluctuations in the atmospheric abundance, since no resurfacing events or other catastrophic processes may occur.

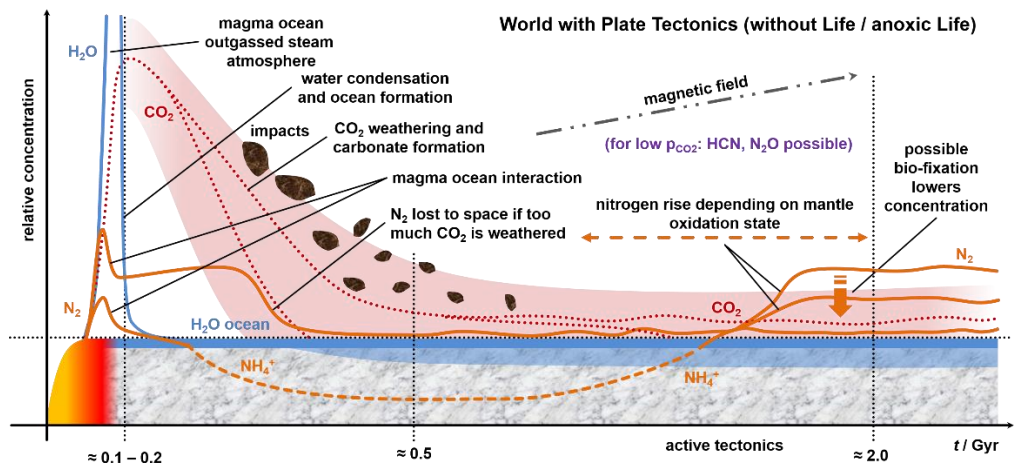

**Fig. 7.** Illustration of the atmospheric evolution of $N_2$ (orange), $CO_2$ (red), and $H_2O$ (blue) for a world where no life (or anoxic life) originated. To show the diversity of the pathways for both $N_2$ and $CO_2$ scenarios (lines) are respectively drawn; for $CO_2$ a range is provided

(red shaded area). $H_2O$ condensed and formed a liquid ocean, $CO_2$ and $N_2$ are weathered out of the atmosphere, plate tectonics operates and transports oxidized material into the lithosphere of the planet so that nitrogen can be efficiently outgassed later via subduction zone volcanos in the form of $N_2$. The nitrogen partial pressure in long-term depends on the abiotic fixation efficiency. With anoxic life, denitrification does not operate under these assumptions and the buildup of a dense $N_2$-dominated atmosphere is unlikely.

## 4.3 Oxic tectonic (Earth-analogue) world

Here, we describe a world with both plate tectonics and life-forms, i.e. a planet like the Earth. We proceed on the assumption that life arose some 3.7 to 4.0 Gyr ago (<0.9 Gyr after the planet's formation), and has contributed to the development of the surface and the atmosphere ever since.

On Earth, we know that plate tectonics is a fundamental process, required not only to gradually oxidize the lithosphere (Section 2.2), but also influencing diverse outgassing rates while recycling sediments and further fractionating the crust (e.g. Elkins-Tanton, 2005; Hacker *et al.*, 2011). Beginning with plume-driven circulation, this phenomenon may be closely tied to life, since it provides fundamental chemical preconditions for the atmosphere such as the composition and therefore the oxidation state (e.g. Höning *et al.*, 2014; Höning and Spohn, 2016; Lee *et al.*, 2016).

In the very beginning, such a planet develops similarly to the anoxic world scenario, but after the origin of life, the influence of biotic processes begins to set in.

In the post steam atmospheric period, a fraction of the $CO_2$ can remain in the atmosphere as a greenhouse gas and thermospheric IR-cooler, inhibiting escape to space and keeping the surface temperature above freezing. Later, the higher solar luminosity and additional greenhouse gases (i.e., $N_2O$, $H_2O$, $CH_4$) partially take over this task. In this period, outgassed (or residual) $N_2$ is not only fixed abiotically, it is also reduced by biotic fixation (e.g. Kharecha *et al.*, 2005; Haqq-Misra *et al.*, 2008; Zerkle and Mikhail, 2017), as described in Section 2.3. However, the relatively low nitrogen partial pressure in combination with the difficulty of re-introducing already sedimented nitrogen into the anoxic nitrogen cycle, constrained the further development of life. Thus, an equilibrium between available (therefore outgassed) nitrogen and the number of lifeforms, which can process and therefore fix it, is established, particularly if one considers a possible uptake of abiotically fixed nitrogen within the oxygen-deficient oceans. The fixed nitrogen is stored in the biosphere and in sediments, as long as no recycling occurs. The above ideas are graphically portrayed in Fig. 8a.

Between 2.7 and 2.2 Gyr ago (~2 Gyr after the planet's formation), not only plate tectonics changes fundamentally to become the process that is observed on the Earth today, but also the atmosphere changes dramatically and the chemical cycles rearrange (e.g. Condie and O'Neill, 2010; Catling, 2014; Zerkle *et al.*, 2017). This transition, known as the GOE, provides oxygen as a nutrient in biological processes and leads to nitrogen-releasing processes such as denitrification (see Fig. 2 and Section 2.2 and references therein). However, for bacteria that are suited to a reducing environment, this implies an existential crisis requiring adaption for survival (e.g. Schopf, 2014; Schirrmeister *et al.*, 2015). Via weathering and uptake processes, sedimented nitrogen can also be cycled back into the ocean and atmosphere, boosting the accumulation of atmospheric $N_2$ (and $N_2O$). At the end of the oxygenation time (see Catling *et al.*, 2005), further oxygen sinks are filled up followed by a second, but smaller rise in $O_2$ partial pressures. A new equilibrium between available and fixed nitrogen is established. At later stages, the $N_2$ abundance could be quite constant as was the case for the last 600 Myr on Earth (Berner, 2006). Finally, the atmosphere is predominantly composed of nitrogen and oxygen, where the latter is significantly less abundant though still being an important constituent (0.78 bar and 0.21 bar on Earth).

## 4 . 4 Entirely extinct world

In the fourth scenario we investigate an Earth-like evolution, but assume that all lifeforms become extinct in the (far) future (illustrated in Fig. 8). In this case, the earlier atmospheric evolution during the first few hundred Myr follows the mechanisms with active plate tectonics as discussed in the previous chapter.

Cyanobacteria, phytoplankton, etc. have evolved permitting a massive recycling of the secondary outgassed atmospheric $N_2$ by biological activity due to denitrification, anammox and anaerobic ammonium oxidation (e.g. Galloway, 2003; Cartigny and Marty, 2013, see also the previous section).

As long as the planet is populated by life forms which efficiently cycle $N_2$ back into the atmosphere, it will be $N_2$-dominated. However, when life becomes extinct, denitrification and anammox stop. The atmosphere no longer experiences strong return fluxes of nitrogen and $N_2$ and $O_2$ become fixed as $NO_x$ primarily by lightning. Then nitrogen is almost completely sequestered into the surface and dissolved into the ocean within about 100 Myr (e.g. Lovelock and Margulis, 1974; Jacob, 1999; Galloway, 2003; Cartigny and Marty, 2013; Lammer *et al.*, 2018). After the life-related oxygen disappears, $N_2$ continues to be fixed further in which the required oxygen is supplied by $H_2O$ or $CO_2$. Similar to the two other cases discussed before, the atmosphere will then, ignoring possible argon, evolve into a $CO_2$-dominated atmosphere (Lovelock and Margulis, 1974; Margulis and Lovelock, 1974).

A significant release of nitrogen at mid-ocean ridges (Kasting *et al.*, 1993), also discussed in Section 4.2, is even more uncertain considering the linkage between a biosphere and plate tectonics (Höning *et al.*, 2014; Höning and Spohn, 2016). Keeping this in mind, an alteration in modern-style plate tectonics is feasible after life ceases to exist. Further, if plate tectonics stops, a condition similar to the stagnant-lid regime world could evolve. Generally, this implies that only the simultaneous dominant presence of N₂ and O₂ in the atmosphere represents a geo-biosignature (see also Stüeken *et al.*, 2016a).

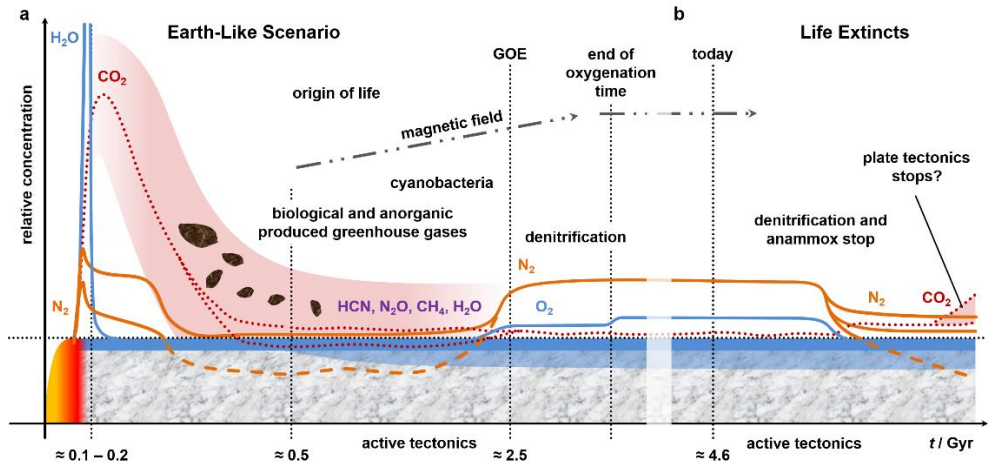

**Fig. 8.** Illustration of the atmospheric evolution of N₂ (orange), CO₂ (red), O₂ (light blue) and H₂O (deep blue) for an Earth-analogue world from its origins to a possible future. To show the diversity of the pathways for both N₂ and CO₂ scenarios (lines) are respectively drawn; for CO₂ a range is provided (red shaded area). a) Due to plate tectonics and liquid water, atmospheric CO₂ will be weathered from the atmosphere and transformed into carbonates. A fraction of the CO₂ remains in the atmosphere. As soon as the planet's crust and upper mantle oxidizes, N₂ is released efficiently via volcanos. After the origin of life, the oxidation state is enhanced with a strong buildup of O₂ during the Great Oxidation Event (GOE). Since then, O₂ has become a major constituent of the atmosphere and lifeforms began to balance the N₂ surface partial pressure to the present level. b) When life becomes extinct and denitrification stops, the atmospheric nitrogen is weathered into the surface within a few tens of Myr, so that the atmosphere will evolve into a thin CO₂- or/and N$_2$-dominated atmosphere, similar to that of Section 4.2. In case that plate tectonics stops, CO$_2$ rises again, leading to a final composition comparable to that of Section 0.

## 5. EARTH-LIKE PLANETS IN HABITABLE ZONES OF M- AND K-TYPE STARS

The atmospheric evolution scenarios discussed above were for planets orbiting solar-like G-type stars. Here we briefly discuss similar scenarios for Earth-like planets with or without active plate tectonics which evolve inside habitable zones around active young M- and K-stars. Atmospheres of such planets, situated in their respective habitable zone around a M or K-type star, would be subject to extreme X-ray and EUV fluxes for a much longer time than if they were orbiting a G-type star (e.g. Gershberg *et al.*, 1999; Scalo *et al.*, 2007; Loyd *et al.*, 2016; Youngblood *et al.*, 2016). Furthermore, dense stellar plasma fluxes ejected by coronal mass ejections (CMEs) (Khodachenko *et al.*, 2007; Lammer *et al.*, 2007; 2009) can initiate various atmospheric escape processes. This affects the planet's habitability in terms of surface water inventory, atmospheric pressure, greenhouse warming efficiency, and the dosage of the UV surface irradiation (e.g. Lammer *et al.*, 2007; Scalo *et al.*, 2007; Luger *et al.*, 2015; Airapetian *et al.*, 2017b).

Besides the exposure of high EUV fluxes and dense stellar plasma, one can expect that Earth-like planets inside the habitable zone of these stars are either partially or totally tidally-locked, resulting in smaller magnetospheres relative to that of the Earth's (our Fig 9; Khodachenko *et al.*, 2007). A recent model by Kislyakova et al. (2017; 2018) suggested that some M-star habitable zone planets could be strongly affected by electromagnetic induction heating during their early evolution, caused by the star's rotation and the planet's orbital motion. In such a case, induction heating can melt the planetary mantle hence inducing extreme volcanic activity and constant resurfacing events, similar to Jupiter's moon Io, though this effect would be somewhat smaller for habitable-zone planets.

Moreover, the atmospheres and oceans of tidally locked planets may freeze out to form a permanent icecap on the dark side of the planet. According to Joshi (2003), $CO_2$-dominated atmospheres with pressures of about 100 mbar may be sustained on tidally locked Earth-like planets through circulation between the day and night sides. In addition, that study suggested that thicker $CO_2$ atmospheres of about $1 - 2$ bar would allow for liquid water on the planet's surface, which was also confirmed by Shields et al. (2013), who assumed such planets in the outer edge of M-star habitable zones. Further, cloud feedback could expand the habitable zone of tidally locked planets significantly (Yang *et al.*, 2013). The formation of a biosphere requires the presence of a stable atmosphere and water inventory. However, various previous studies (Khodachenko *et al.*, 2007; Lammer *et al.*, 2007; 2013; Airapetian *et al.*, 2017b) indicate that Earth-like planets inside the habitable zones of M-stars most likely do not build up dense atmospheres over long time periods, due to thermal and nonthermal atmospheric escape processes, geophysical difficulties related to plate tectonics (Lammer *et al.*, 2009), and surface weathering of $CO_2$.

Planets located around M- and K-stars could catastrophically outgas dense steam atmospheres during the solidification of their magma oceans (e.g. Elkins-Tanton, 2012). Similar to G-star planets, after about $1 – 2$ Myr the water vapor then condenses (Hamano *et al.*, 2013; Lebrun *et al.*, 2013; Massol *et al.*, 2016; Salvador *et al.*, 2017) and a liquid ocean forms. $CO_2$ would then weather out of the atmosphere (e.g. Walker *et al.*, 1981; Alt and Teagle, 1999; Lammer *et al.*, 2018 and references therein).

Lammer et al. (2007) and Airapetian et al. (2017b) suggested that nonthermal atmospheric ion escape caused by radiative forcing could incur a significant atmospheric loss rate due to the long lasting high stellar EUV flux of M- and K-type stars. Fig. 9b shows the response of the exobase level for an Earth-type nitrogen atmosphere in relation to the incident EUV flux. For young planets, the EUV-heated upper atmosphere exceeds the magnetopause levels easily leading to massive nitrogen loss rates of 300 mbar/Myr for 10 $F_{EUV,Earth}$ and much more for higher fluxes (Lichtenegger *et al.*, 2010). In the case of $CO_2$-atmospheres, $CO_2$ can be massively dissociated, but the products (oxygen, carbon) cannot accumulate due to efficient escape, thus an Earth-size planet is not able to build up a dense atmosphere (Lammer *et al.*, 2007; Tian, 2009; Airapetian *et al.*, 2017b). Moreover, Airapetian et al. (2017b) found that the escape time of a 1 bar atmosphere on a terrestrial-type planet in the habitable zone of Proxima Cen b is expected to be about 10 Myr. One can conclude that atmospheres of M- and K-star Earth-like habitable zone planets, which could maintain liquid water oceans on their surface, are unlikely to build up $N_2$-dominated atmospheres. Due to high thermal and non-thermal atmospheric escape rates, the remaining thin atmospheres will most likely be $CO_2$-dominated similar to present Mars, although further (model) studies are needed.

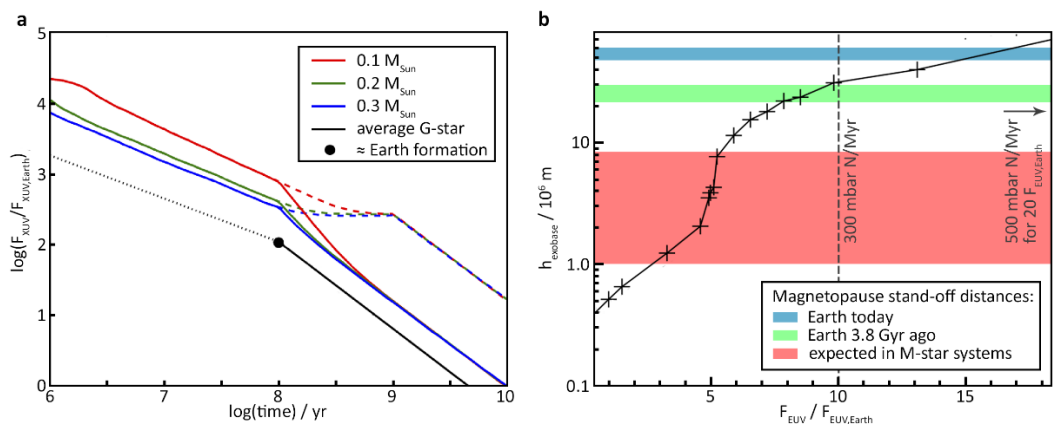

**Fig. 9.** a) Incident XUV flux over time for planets in the habitable zone of M-type stars and of a moderate rotating G-type star, normalized to the Earth's present flux. Scenarios for an EUV saturation time of 0.1 Gyr are drawn solid whereby dashed lines correspond to 1.0 Gyr. Figure adapted from Luger et al. (2015). b) Variation of the exobase altitude relative to the Earth-normalized incident EUV flux after Tian et al. (2008a) in relation to planetary magnetopause stand-off distances under different stellar wind conditions (data for M-type stars: Khodachenko *et al.*, 2007; data for Earth: Lichtenegger *et al.*, 2010). A nitrogen escape rate of 300 mbar/Myr is expected for 10 $F_{EUV,Earth}$ respectively 500 mbar/Myr for 20 $F_{EUV,Earth}$ (Lichtenegger *et al.*, 2010).

## 6.   IMPLICATIONS FOR THE SEARCH OF LIFE ON EARTH-LIKE EXOPLANETS AND POSSIBLE DETECTION METHODS

The central issues regarding the existence of a potential Earth-like biosphere on a hypothetical ocean-surface environment of an extraterrestrial Earth-like planet are connected to the oxidation states of atmosphere and interior, the need for a fully oxidized surface und uppermost mantle and the likely necessity of plate tectonics. Therefore, the buildup of a dense $N_2$-dominated atmosphere is strongly linked to the planetary oxygenation time, atmospheric $O_2$ and lifeforms that are capable of denitrification (Zerkle and Mikhail, 2017; Zerkle *et al.*, 2017; Lammer *et al.*, 2018 and references therein).

Plate tectonics is a crucial factor for maintaining the activity of the carbon-silicate cycle over geological timescales (e.g. Walker *et al.*, 1981; Kasting, 1993; McCammon, 2005; Southam *et al.*, 2015; Krissansen-Totton *et al.*, 2018a). The manner in which plate tectonics starts and operates, and whether it is geophysically stable (or transient) over the planetary lifetime, are however not fully understood (e.g. Tackley, 2000; Bercovici, 2003; Bercovici and Ricard, 2003; van Hunen and Moyen, 2012; Gerya *et al.*, 2015; O'Neill *et al.*, 2016). Such feedbacks are likely necessary for determining the outgassing of $N_2$-dominated atmospheres, which is related to the origin of complex life and long term habitability.

Water-rich Earth-like planets inside the habitable zone without both active plate tectonics and lifeforms which are capable of denitrification, most likely never build up dense secondary outgassed $N_2$-dominated atmospheres. One should keep in mind that on ocean worlds, whether life exists there or not (Noack *et al.*, 2016), the buildup of $CO_2$-dominated atmospheres is unlikely due to proposed destabilizing climate feedbacks (Kitzmann *et al.*, 2015), but we do not focus on such scenarios here. In the case of M- and K-type host stars, due to the stars' long-lasting active radiation and plasma environment, the atmospheres of Earth-like planets inside the habitable zones of these stars will experience high mass loss rates, which could also prevent the formation of dense secondary atmospheres.

Especially if the planet lacks a magnetosphere then thin atmospheres (surface partial pressures of a few mbar to a few tens of mbar) will not greatly protect the surface environment from the exposure to highly energetic cosmic rays (Brack *et al.*, 2010; Belisheva *et al.*, 2012; Grießmeier *et al.*, 2016). In that case, secondary radiation caused by particle showers in such a thin atmosphere will likely have global effects and may prevent life or sterilize the planet's surface (e.g. Belisheva *et al.*, 1994; Belisheva and Popov, 1995; Belisheva and Emelin, 1998; Dar *et al.*, 1998; Belisheva and Gak, 2002; Smith and Scalo, 2004;

Grießmeier *et al.*, 2005; Brack *et al.*, 2010; Pavlov *et al.*, 2012). In the case of an Earth-like planet, the effects of highly energetic particles on biological systems could be strongly reduced or negligible because life forms would keep the atmosphere dense enough through denitrification such that high energy particles are shielded from the surface in an efficient way.

Moreover, according to Catling et al. (2005), the rather long oxygenation time could preclude complex life on Earth-like planets orbiting early-type stars, which end their main sequence lives before sufficient oxygenation can occur. Conversely, Earth-like planets inside the habitable zones of solar-like G-stars are potentially more favorable habitats for the evolution of complex life forms.

To conclude, in accordance with our argumentation and in agreement with Stüeken et al. (2016a), we expect $N_2$ and $O_2$ as major constituents of terrestrial planetary atmospheres to be a geo-biosignature for a biosphere populated by highly developed life forms. The bacterial by-product, $N_2O$, provides a similar indication (Muller, 2013).

Among the many instruments and space missions currently under development, PLATO (Rauer *et al.*, 2014) is in the best position for the detection of transiting Earth-size planets orbiting in the habitable zone of G-type stars. Due to their transiting geometries and bright central host stars, the atmosphere of these planets can then be observed and characterized through multi-wavelength transmission spectroscopy (e.g. Seager *et al.*, 2000; Brown, 2001). The coronagraphs on-board future space missions currently under development, such as HABEX (Mennesson *et al.*, 2016) and LUVOIR (Bolcar *et al.*, 2016), will have the capability to directly image low-mass planets in the habitable zone of G-type stars and directly measure their emission spectra. For transiting planets, future large missions may allow us to additionally obtain transmission spectra. However, these may not be able to probe the deepest atmospheric layers where most of the water resides, due to refraction (García Muñoz *et al.*, 2012; Misra *et al.*, 2014b). Fujii et al. (2018) provide a comprehensive review of future planned observations in a biosignature context.

In the previous sections we showed that from our current knowledge of the Earth's atmospheric evolution, it is possible to conclude that the detection of an $N_2$-dominated atmosphere presenting a strong component of $O_2$ and a negligible amount of $CO_2$, possibly accompanied by the presence of $O_3$, $H_2O$, $CH_4$ and $N_2O$, can decisively indicate the presence of an Earth-like habitat and therefore an aerobic biosphere (Airapetian *et al.*, 2017a). The possible detectability of an anoxic habitat, such as that of Earth during the Archean with a $CH_4$-rich atmosphere, is given by Arney et al. (2016) and Krissansen-Totton et al. (2018b). Oxygen and water molecules feature several transition bands ranging from the near-UV to the infrared, some of them particularly strong, that can be used to detect and measure the abundance of those molecules. Similarly, $CO_2$ presents strong molecular transition bands in the infrared, which can thus be used to detect and measure its abundance (e.g. Rauer *et al.*, 2011; Hedelt *et al.*, 2013; Bétrémieux and Kaltenegger, 2013; Arney *et al.*, 2016), for example through transmission spectroscopy carried out with high-resolution spectrographs attached to the Extremely Large Telescope (ELT) (e.g. de Kok *et al.*, 2013; Brogi *et al.*, 2014; Snellen *et al.*, 2017).

The detection of nitrogen is far more challenging. However, nitrogen oxides are deeply linked to biological and atmospheric processes (Muller, 2013). For instance, on Earth $N_2O$ would be efficiently depleted by photodissociation in the troposphere, if it were not protected by the ozone layer. $NO_2$ and $N_2O$ feature molecular transition bands in the blue optical region and at near-UV wavelengths, but their strength is significantly smaller than those of other molecules (e.g., $O_2$, $O_3$) located at similar wavelengths (Bétrémieux and Kaltenegger, 2013), making their detection challenging. $N_2O$ has two further infrared absorption bands (at about 4.5 and 7.8 µm), although these are generally also rather weak (Rauer *et al.*, 2011; Hedelt *et al.*, 2013). Pallé et al. (2009) showed that the infrared absorption features of the $O_2 \cdot O_2$ and $O_2 \cdot N_2$ dimers at 1.26 µm are in principle detectable in the transmission spectrum of an Earth-like planet, assuming the spectral resolution and signal-to-noise ratio are high enough. Misra et al. (2014a) showed that the analysis of the dimer $O_2 \cdot O_2$ may lead to a measurement of the atmospheric $O_2$ pressure, which would be extremely valuable for the characterization of Earth-like planets. However, a similar study has not yet been carried out for the $O_2 \cdot N_2$ dimer, which may lead to the detection of nitrogen and, more importantly, to a measurement of the $N_2$ atmospheric abundance and pressure. Observationally, the $O_2 \cdot N_2$ dimer could be detected and measured from the ground using the ELTs.

The upper atmosphere of the Earth is mostly composed of atomic hydrogen, nitrogen, and oxygen. Various processes, such as resonance scattering and charge exchange with the solar wind lead to the formation of a number of emission lines of atomic H, N, and O, the strongest ones located between X-rays and optical wavelengths. These atoms, located in the upper atmosphere, would also appear in transmission spectra of transiting Earth-like planets. In this case, due to the low atmospheric density, the most favorable features are those in the far-UV (122-200 nm). The reason is that the relatively weak stellar far-UV fluxes would be partially compensated by the extended upper planetary atmosphere at these wavelengths. If strong enough, these emission and/or absorption features might be detectable, allowing to reveal indirectly the presence or absence of a given element in the atmosphere of an Earth-like exoplanet. To this end, these lines might be detectable by future large aperture multi-purpose telescopes, such as LUVOIR, thanks to the unprecedented high far-UV sensitivity of the LUMOS (France *et al.*, 2017) and POLLUX spectrographs (Bouret *et al.*, 2018).

Finally, an $N_2$-$O_2$-atmosphere is transparent for sunlight due to the oxidation of visible-light absorbers through $O_2$. This allows effective Rayleigh scattering on $O_2$-molecules leading to the famous "blue planet"-appearance (Krissansen-Totton *et al.*, 2016b), which makes an Earth-like Rayleigh absorption feature in the visible consistent with an $N_2$-$O_2$ atmosphere a biosignature.

Given the great opportunity that the detection of atmospheric nitrogen compounds would provide for the identification of Earth-like habitats, it is important for future studies to address thoroughly the detectability of this species in the atmosphere of Earth-like planets and to establish what kind of information (such as detection alone, the abundances, pressures, etc.) a given method could provide. It is also important to note that it is crucial to study the detectability of these features for planets with atmospheric pressures and compositions slightly different from those of the Earth and also orbiting stars different from the Sun.

## 7. CONCLUSION

There is a strong correlation between an $N_2$-dominated atmosphere and other constituents like $O_2$, $O_3$, $H_2O$. Such dual detections constitute a biosignature for aerobe life. The latter is the vital contributor to maintain nitrogen dominated atmospheres and drastically impact the composition of an Earth-like atmosphere via complex interactions. Plate tectonics remain another crucial factor and the possible influence of life to its development is yet to be thoroughly investigated. Conversely, it is likely that that life on the Earth would have developed differently, if plate tectonic processes had not operated efficiently. Thus, the detection of atmospheric nitrogen would indicate a tectonically active world ("geo-signature"), whereas $N_2$ and $O_2$ in combination represent a geo-biosignature, while we do not rule out that an anaerobic biosphere is present without maintaining such conditions.

If such an interplay of atmosphere, lithosphere and biosphere providing a base for highly developed life is rare, the atmospheres of most terrestrial planets in the habitable zones would be $CO_2$-dominated. This molecule presents a number of absorption bands in the infrared, making it detectable from the ground with high-resolution spectrographs attached to the ELTs. Our hypothesis could therefore be proven by characterizing the atmosphere of the Earth-size planets detected by TESS and PLATO. The next step could then be the quantification of the habitability occurrence rate, which would require the detection of nitrogen, thus the use of the next generation ground- and space-based telescopes (i.e., ELTs, LUVOIR).

The buildup of any substantial amount of nitrogen might be still more difficult around M- and also K-stars, due to both different geodynamics and their long-lasting phase of strong EUV irradiation leading to severe escape.

Identifying nitrogen and oxygen as primary ingredients of an exoplanet's atmosphere is not only an indication for an oxidized lithosphere and a tectonically active world, it also indicates the existence of an aerobe biosphere.

## APPENDIX

To give an overview on the estimations of nitrogen exchange rates within the Earth's nitrogen cycle of today, some tables (Tab. 2, Tab. 3, Tab. 4, and Tab.5) are attached here. These tables contain the respective outgoing rates of the reservoirs atmosphere (with atmospheric fixed), soil (with land biota), ocean (with ocean biota), and lithosphere. A visualization of this simplified description is given in Fig. 10. All total rates, given in the lower part of the tables, refer to the exchanges between the four combined complexes, readily identifiable in the graphic.

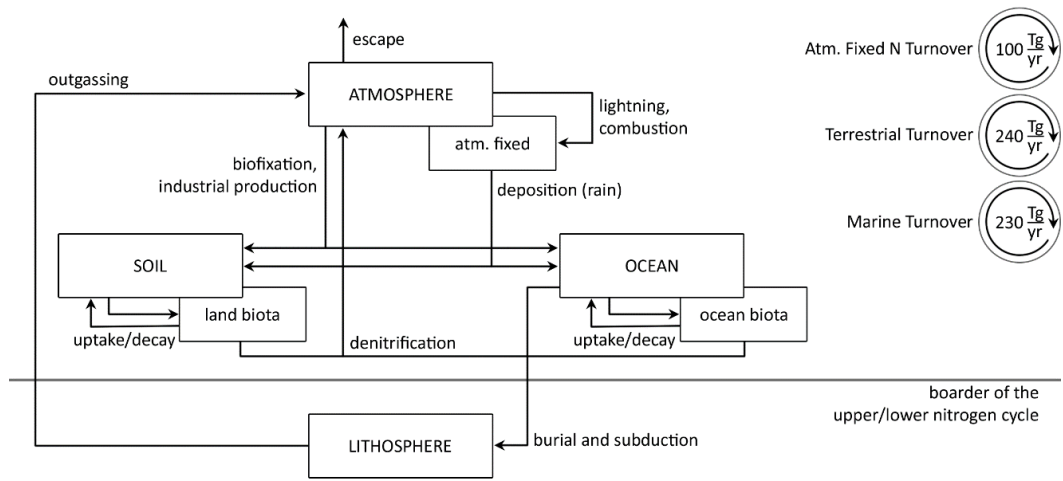

**Fig. 10.** Simplification of the Earth's present day nitrogen cycle. There are four main reservoirs (atmosphere, soil, ocean, and lithosphere), which build the basic structure for the different rates in Tab. 2, Tab. 3, Tab. 4, and Tab. 5.

Note that the yearly turnovers (see also Fig. 10) are high in respect to the total global nitrogen exchange rates between the different reservoirs, which means, that nitrogen is not only processed in a global way, but rather also efficiently (re)cycled in the smaller sub-systems.

**Tab. 2.** Significant nitrogen depletion rates from the atmosphere in (Tg N)/yr.

| Process | Receiving Reservoir | [*Jaffe*, 1992] | [*Jacob*, 1999] | [*Galloway*, 2003] [a] | [*Fowler* et al., 2013] | Diverse Sources |
|---|---|---|---|---|---|---|
| atmosphere | | | | | | |
| industrial production | soil, l. bio. | 40 (40) | 80 (80) | 79 (79) | 120 (120) | 60 (60) [f] |
|   share of fertilizer prod. | l. bio. | | 80 (80) | | 100 (100) | |
| fossil fuel combustion | atm. f. | 20 (20) | | 23 (23) | 30 (30) | |
| biomass burning (fixation) [b] | atm. f. | 12 (10) [c] | 25 (−) [d] | | | |
| biofixation (land) | l. bio. | 150 (−) | 160 (−) | 146 (41) | 118 (60) | 110 (−) [g] |
|   agricultural percentage | l. bio. | | | 41 (41) | 60 (60) | |
|   naturogenic percentage | l. bio. | | | 105 (0) | 58 (0) | |
| biofixation (ocean) | o. bio. | 40 (0) | 20 (−) | 68 (0) | 140 (0) | 40 (−) [g] |
| lightning | atm. f. | 5 (0) | 5 (0) [d] | 4 (0) | 5 (0) [e] | 3 (0) [g] |
| loss to the stratosphere | - | | 9 (0) | | | 9 (0) [g] |
| atmosphere: fixed N | | | | | | |
| deposition (land) | soil | 125 (0) | 80 (0) | 74 (0) | 74 (0) [e] | |
| deposition (ocean) | ocean | 34 (0) | 30 (0) | 30 (0) | 31 (0) [e] | 76 (0) [h] |
|   Total Rate | | | | | | |
| to soil / land biota | | 315 | 320 | 299 | 312 | |
| to ocean / ocean biota | | 74 | 50 | 98 | 171 | |
| to lithosphere | | 0 | 0 | 0 | 0 | |

a: Mean of values given by [*Galloway* et al., 1995] and [*Schlesinger*, 1997], summarized in [*Galloway*, 2003, p. 567].
b: Here, biomass burning represents primarily nitrogen fixation of atmospheric dinitrogen, similar to fossil fuel combustion. In some models, biomass burning describes volatilization of fuels' organic nitrogen during the burning process. Therefore, there is also biomass burning in the land biota outgoing flux table.
c: The anthropogenic percentage is estimated, because there are only explanations and no reliable values given in the original description.
d: Originally, there is a flux "combustion and lightning" of 30 Tg N/yr, whereby lightning is declared as minor part. The splitting into biomass burning (including fossil fuel combustion) and lightning is merely estimated.
e: In the original study, lightning products are declared as directly washed out into soil. Therefore, it is proportionally added to deposition rates in this table.
f: [*Rosswall*, 1983], g: [*Stedman and Shetter*, 1983], h: [*Voss* et al., 2013]

**Tab. 3.** Significant nitrogen depletion rates from the soil in (Tg N)/yr.

| Process | Receiving Reservoir | [*Jaffe*, 1992] | [*Jacob*, 1999] | [*Galloway*, 2003] [a] | [*Fowler* et al., 2013] | Diverse Sources |
|---|---|---|---|---|---|---|
| soil | | | | | | |
| leaching and river runoff | ocean | 34 (0) | 40 (0) | 76 (0) | 80 (0) | |
|   groundwater percentage | ocean | | | | 4 (0) | |
|   riverine flux [b] | ocean | | | | 40 − 70 (0) | |
| biotic uptake | l. bio. | | 2300 (2300) | | | |
| land biota | | | | | | |
| biomass burning (volit.) [c] | atm. f. | | | 19 (19) | 5 (4) | |
| soil/agriculture emission [d] | atm. f. | 130 (> 5) | 80 (−) | 60 (−) | 65 (41) | |
|   $NH_3$ percentage | atm. f. | 122 (−) | | 60 (−) | 60 (40) | |
|   NO percentage | atm. f. | 15 (0) | | | 5 (1) | |
| burial / decay | soil | | 2500 (2500) | | | |
| denitrification | atm. | 147 (0) | 130 (0) | 177 (0) | 113 (7) | 125 (−) [g] |
|   $N_2O$ percentage | atm. | | | 10 (0) | 13 (7) | |
|   Total Rate | | | | | | |
| to atmosphere / atm. fixed | | 284 | 210 | 256 | 183 | |
| to ocean / ocean biota | | 34 | 40 | 76 | 80 | |
| to lithosphere | | 0 | 0 | 0 | 0 | |

a: Mean of values given by [*Galloway* et al., 1995] and [*Schlesinger*, 1997], summarized in [*Galloway*, 2003, p. 567].
b: Fish landing is often not mentioned or, if the level of detail is high enough, already subtracted from river runoff. Voss et al. [2013] estimated a flux of 3.7 (Tg N)/yr, which would be clearly large enough to be considerable.
c: This represents primarily the nitrogen within the fuel which is volatilized, thus, in this case, this flux is not a source for fixed nitrogen. In some models the biomass burning flux describes a fixing process of atmospheric $N_2$, similar to fossil fuel combustion. Therefore, there is also biomass burning in the atmospheric outgoing flux table.
d: Denitrification rates are indeed also a form of soil emission. Here they are listed separately for a better understanding.
g: [*Stedman and Shetter*, 1983]

**Tab. 4.** Significant nitrogen depletion rates from the ocean in (Tg N)/yr.

| Process | Receiving Reservoir | [*Jaffe*, 1992] | [*Jacob*, 1999] | [*Galloway*, 2003] [a] | [*Fowler* et al., 2013] | Diverse Sources |
|---|---|---|---|---|---|---|
| ocean | | | | | | |
|   ocean emission | atm., atm. f. | | ≈ 0 (0) | 13 (0) | 14.5 (0) | |
|     NH₃ percentage | atm. f. | | | 13 (0) | 9 (0) [b] | |
|     N₂O percentage | atm. | | | | 5.5 (3) | |
|   biotic uptake | o. bio. | | 1600 (1600) | | | |
| ocean biota | | | | | | |
|   denitrification | atm. | 30 (0) | 100 (0) | 141 (0) | 100 − 280 (0) | 31 (−) [d] |
|     N₂O percentage | atm. | | | 3 (0) | 5.5 (0) | |
|   burial / decay | ocean | | 1600 (0) | | | |
|   burial and subsidence | lith. | 14 (0) [c] | 10 (0) | ∼0 (0) | 20 (0) | 1.3 [e] |
|     Total Rate | | | | | | |
| to atmosphere / atm. fixed | | 30 | 100 | 154 | 114.5 − 294.5 | |
| to soil / land biota | | 0 | 0 | 0 | 0 | |
| to lithosphere | | 14 | 10 | 0 | 20 | |

a: Mean of values given by [*Galloway* et al., 1995] and [*Schlesinger*, 1997], summarized in [*Galloway*, 2003, p. 567].
b: This value of NH₃ percentage also contains volcanic emissions
c: While most studies lack any entry of weathering processes back from seafloor to the ocean, this study has such a flux. Here, the net value of subducted nitrogen is given.
d: [*Stedman and Shetter*, 1983], e: [*Hilton* et al., 2002]

**Tab. 5.** Significant nitrogen depletion rates from the lithosphere in (Tg N)/yr.

| Process | Receiving Reservoir | [*Jaffe*, 1992] | [*Hilton* et al., 2002] | [*Sano* et al., 2001] | [*Catling and Kasting*, 2017] | Diverse Sources |
|---|---|---|---|---|---|---|
| lithosphere | | | | | | |
|   outgassing: MORs | atm. | 1.000 | 0.0280 | 0.0616 | 0.1064 [b] | 0.0616 [c] |
|   outgassing: island arcs | atm. | | 0.5546 | 0.0179 | 1.4706 | 0.9804 [d] |
|     recycled percentage | atm. | | | 0.0126 | 0.4202 | |
|   outgassing: BABs | atm. | | | 0.0157 | | |
|     recycled percentage | atm. | | | 0.0067 | | |
|   seafloor weathering [a] | ocean | 5.000 | | | | 14.000 [e] |
| lower mantle | | | | | | |
|   hotspot volcanic action | atm. | | | 0.0001 | 0.8403 | |
|     Total Rate | | | | | | |
| to atmosphere / atm. fixed | | 1.000 | 0.5826 | 0.0953 | 2.4173 | |
| to soil / land biota | | 0 | 0 | 0 | 0 | |
| to oceanic / ocean biota | | 5.000 | 0 | 0 | 0 | |

a: Some studies lack this entry and take the "burial and subsidence" from ocean to lithosphere as net flux.
b: Taken from [*Marty* et al., 2013]
c: [*Marty*, 1995], d: [*Fischer*, 2008], e: [*Stedman and Shetter*, 1983]

# BIBLIOGRAPHY

Airapetian, V. S., Glocer, A., Gronoff, G., Hébrard, E., and Danchi, W. (2016) Prebiotic chemistry and atmospheric warming of early Earth by an active young Sun. *Nature Geoscience* 9:452-455

Airapetian, V. S., Jackman, C. H., Mlynczak, M., Danchi, W., and Hunt, L. (2017a) Atmospheric Beacons of Life from Exoplanets Around G and K Stars. *Sci Rep* 7

Airapetian, V. S., Glocer, A., Khazanov, G. V., Loyd, R. O., France, K., Sojka, J., Danchi, W. C., and Liemohn, M. W. (2017b) How Hospitable Are Space Weather Affected Habitable Zones? The Role of Ion Escape. *Astrophys J Lett* 836

Alt, J. C. and Teagle, D. A. (1999) The uptake of carbon during alteration of ocean crust. *Geochim Cosmochim Acta* 63:1527-1535

Arney, G., Domagal-Goldman, S. D., Meadows, V. S., Wolf, E. T., Schwieterman, E., Charnay, B., Claire, M., Hébrard, E., and Trainer, M. G. (2016) The Pale Orange Dot: The Spectrum and Habitability of Hazy Archean Earth. *Astrobiology* 16:873-899

Arney, G. N., Meadows, V. S., Domagal-Goldman, S. D., Deming, D., Robinson, T. D., Tovar, G., Wolf, E. T., and Schwieterman, E. (2017) Pale Orange Dots: The Impact of Organic Haze on the Habitability and Detectability of Earthlike Exoplanets. *Astrophys J* 836

Aulbach, S. and Stagno, V. (2016) Evidence for a reducing Archean ambient mantle and its effects on the carbon cycle. *Geology* 44 (9):751-754

Avice, G., Marty, B., Burgess, R., Hofmann, A., Philippot, P., Zahnle, K., and Zakharov, D. (2018) Evolution of atmospheric xenon and other noble gases inferred from Archean to Paleoproterozoic rocks. *Geochim Cosmochim Acta* 232:82-100

Barry, P. H. and Hilton, D. R. (2016) Release of subducted sedimentary nitrogen throughout Earth's mantle. *Geochem Persp Let* 2:148-159

Belisheva, N. K., Popov, A. N., and Poniavin, D. I. (1994) Biological effects in cell cultures and geomagnetic field. In *Proceedings of the International Symposium on "Charge and Field Effects in Biosystems - 4"* M. J. Allen, S. F. Cleary, and A. E. Sowers, World Scientific Publishing Co. Pte. Ltd., Richmond, VA, pp 159-173

Belisheva, N. K. and Popov, A. N. (1995) Dynamics of the morphofunctional state of cell cultures with variation in the geomagnetic field in high latitudes. *Biophysics* 40:737-745

Belisheva, N. K. and Emelin, C. E. (1998) Self-organisation of living systems under geocosmical agents impact. *Nauchnoe priborostroenie RAS* 35-37

Belisheva, N. K. and Gak, E. Z. (2002) Significance of cosmic rayvariations for biosystem functions. In *Proceedings of the VII International Conference on Ecology and Development of North-West Russia* Saint Petersburg, pp 118-120

Belisheva, N. K., Lammer, H., Biernat, H. K., and Vashenuyk, E. V. (2012) The effect of cosmic rays on biological systems - an investigation during GLE events. *Astrophys Space Sci Trans* 8:7-17

Bercovici, D. (2003) The Generation of plate tectonics from mantle convection. *Earth Planet Sci Lett* 6451:107-121

Bercovici, D. and Ricard, Y. (2003) Energetics of a two-phase model of lithospheric damage, shear localization and plate-boundary formation. *Geophys J Int* 152:581-596

Berner, R. A. (2006) Geological nitrogen cycle and atmospheric N2 over Phanerozoic time. *Geology* 34:413-415

Bétrémieux, Y. and Kaltenegger, L. (2013) Transmission Spectrum of Earth as a Transiting Exoplanet from the Ultraviolet to the Near-infrared. *Astrophys J* 772

Bolcar, M. R., Balasubramanian, K., Crooke, J., Feinberg, L., Quijada, M., Rauscher, B. J., Redding, D., Rioux, N., Shaklan, S., Stahl, H. P., Stahle, C. M., and Thronson, H. (2016) Technology gap assessment for a future large-aperture ultraviolet-optical-infrared space telescope. *J Astron Telesc Instrum Syst* 2

Bouret, J.-C., Neiner, C., Gómez de Castro, A. I., Evans, C., Gaensicke, B., Shore, S., Fossati, L., Gry, C., Charlot, S., Marin, F., Noterdaeme, P., and Chaufray, J.-Y. (2018) The science case for POLLUX: a high-resolution UV spectropolarimeter onboard LUVOIR. *Proceedings of the SPIE, Volume 10699, id. 106993B*

Boyd, S. R. and Philippot, P. (1998) Precambrian ammonium biogeochemistry: a study of the Moine metasediments, Scotland. *Chem Geol* 144:257-268

Boyd, S. R. (2001) Ammonium as a biomarker in Precambrian metasediments. *Precambrian Res* 108:159-173

Brack, A., Horneck, G., Cockell, C. S., Bérces, A., Belisheva, N. K., Eiroa, C., Henning, T., Herbst, T., Kaltenegger, L., Léger, A. *et al.* (2010) Origin and Evolution of Life on Terrestrial Planets. *Astrobiology* 10:69-76

Brogi, M., de Kok, R. J., Birkby, J. L., Schwarz, H., and Snellen, I. A. (2014) Carbon monoxide and water vapor in the atmosphere of the non-transiting exoplanet HD 179949 b. *Astron Astrophys* 565

Brown, T. M. (2001) Transmission Spectra as Diagnostics of Extrasolar Giant Planet Atmospheres. *Astrophys J* 553:1006-1026

Busigny, V. and Bebout, G. E. (2013) Nitrogen in the Silicate Earth: Speciation and Isotopic Behavior during Mineral–Fluid Interactions. *Elements* 9:353-358

Canfield, D. E., Glazer, A. N., and Falkowski, P. G. (2010) The Evolution and Future of Earth's Nitrogen Cycle. *Science* 330:192-196

Cartigny, P., Pineau, F., Aubaud, C., and Javoy, M. (2008) Towards a consistent mantle carbon flux. *Earth Planet Sci Lett* 265:672-685

Cartigny, P. and Marty, B. (2013) Nitrogen Isotopes and Mantle Geodynamics: The Emergence of Life and the Atmosphere-Crust-Mantle Connection. *Elements* 9:359-366

Catling, D. C. (2014) The Great Oxidation Event Transition. In *About Reference Module in Earth Systems and Environmental Sciences* 6, S. A. Elias, Elsevier Inc., pp 177-195

Catling, D. C., Zahnle, K. J., and McKay, C. P. (2001) Biogenic Methane, Hydrogen Escape, and the Irreversible Oxidation of Early Earth. *Science* 293:839-843

Catling, D. C. and Claire, M. W. (2005) How Earth's atmosphere evolved to an oxic state: A status report. *Earth Planet Sci Lett* 237

Catling, D. C., Glein, C. R., Zahnle, K. J., and McKay, C. P. (2005) Why O2 Is Required by Complex Life on Habitable Planets and the Concept of Planetary "Oxygenation Time". *Astrobiology* 5:415-438

Catling, D. C. and Kasting, J. F. (2017), Atmospheric Evolution on Inhabited and Lifeless Worlds Cambridge University Press

Chameides, W. L. and Walker, J. C. (1981) Rates of Fixation by Lightning of Carbon and Nitrogen in Possible Primitive Atmospheres. *Origins of Life* 11:291-302

Charnay, B., Le Hir, G., Fluteau, F., Forget, F., and Catling, D. C. (2017) A warm or a cold early Earth? New insights from a 3-D climate-carbon model. *Earth Plan Sci Lett* 474:97-109

Chyba, C. F. and Sagan, C. (1992) Endogenous production, exogenous delivery and impact-shock synthesis of organic molecules: an inventory for the origins of life. *Nature* 335:125-132

Cockell, C. S., Kaltenegger, L., and Raven, J. A. (2009) Cryptic Photosynthesis-Extrasolar Planetary Oxygen Without a Surface Biological Signature. *Astrobiology* 9:623-636

Coltice, N., Simon, L., and Lécuyer, C. (2004) Carbon isotope cycle and mantle structure. *Geophys Res Lett* 31

Condie, K. C. and O'Neill, C. (2010) The Archean-Proterozoic Boundary: 500 My of Tectonic Transition in Earth History. *American Journal of Science* 310:775-790

Coogan, L. A. and Gillis, K. M. (2018) Low-Temperature Alteration of the Seafloor: Impacts on Ocean Chemistry. *Annu Rev Earth Planet Sci* 46:21-45

Cooray, V. (2015) Interaction of Lightning Flashes with the Earth's Atmosphere. In *An Introduction to Lightning* Springer, Dordrecht, pp 341-358

Coustenis, A. and Taylor, F. W. (2008) Titan: Exploring an Earthlike World. Second Edition. In *Series on Atmospheric, Oceanic and Planetary Physics, Vol 4* A. Coustenis and F. W. Taylor, World Scientific Publishing Co, Singapore, pp 154-155

Dar, A., Laor, A., and Shaviv, N. J. (1998) Life Extinctions by Cosmic Ray Jets. *Phys Rev Lett* 80:5813-5816

de Kok, R. J., Brogi, M., Snellen, I. A., Birkby, J., Albrecht, S., and de Mooij, E. J. (2013) Detection of carbon monoxide in the high-resolution day-side spectrum of the exoplanet HD 189733b. *Astron Astrophys* 554

Delano, J. W. (2001) Redox history of the Earth's interior since approximately 3900 Ma: implications for prebiotic molecules. *Orig Life Evol Biosph* 31:311-341

Des Marais, D. J., Harwit, M. O., Jucks, K. W., Kasting, J. F., Lin, D. N., Lunine, J. I., Schneider, J., Seager, S., Traub, W. A., and Woolf, N. J. (2002) Remote Sensing of Planetary Properties and Biosignatures on Extrasolar Terrestrial Planets. *Astrobiology* 2:153-181

Devol, A. H. (2003) Nitrogen cycle: Solution to a marine mystery. *Nature* 422:575-576

Domagal-Goldman, S. D., Meadows, V. S., Claire, M. W., and Kasting, J. F. (2011) Using Biogenic Sulfur Gases as Remotely Detectable Biosignatures on Anoxic Planets. *Astrobiology* 11:419-441

Domagal-Goldman, S. D., Segura, A., Claire, M. W., Robinson, T. D., and Meadows, V. S. (2014) Abiotic Ozone and Oxygen in Atmospheres Similar to Prebiotic Earth. *Astrophys J* 792

Elkins-Tanton, L. T. (2008) Linked magma ocean solidification and atmospheric growth for Earth and Mars. *Earth Planet Sci Lett* 271:181-191

Elkins-Tanton, L. T. (2005) Continental magmatism caused by lithospheric delamination. *Geological Society of America, Spcial Papers* 388:449-461

Elkins-Tanton, L. T. (2012) Magma Oceans in the Inner Solar System. *Annu Rev Earth Planet Sci* 40:113-139

Fegley, B. J., Prinn, R. G., Hartman, H., and Watkins, G. H. (1986) Chemical effects of large impacts on the earth's primitive atmosphere. *Nature* 319:305-308

Fischer, T. P. (2008) Fluxes of volatiles (H2O, CO2, N2, Cl, F) from arc volcanoes. *Geochemical Journal* 42:21-38

Fischer, T. P. (2008) Fluxes of volatiles (H2O, CO2, N2, Cl, F) from arc volcanoes. *Geochemical Journal* 42:21-38

Fossati, L., Erkaev, N. V., Lammer, H., Cubillos, P. E., Odert, P., Juvan, I., Kislyakova, K. G., Lendl, M., Kubyshkina, D., and Bauer, S. J. (2017) Aeronomical constraints to the minimum mass and maximum radius of hot low-mass planets. *Astron Astrophys* 598

Fowler, D., Coyle, M., Skiba, U., Sutton, M. A., Cape, J. N., Reis, S., and Sheppard, L. J. (2013) The global nitrogen cycle in the twenty-first century. *Philosophical Transactions Royal Society B 368*

France, K., Fleming, B., West, G., McCandliss, S. R., Bolcar, M. R., Harris, W., Moustakas, L., O'Meara, J. M., Pascucci, I., Rigby, J., Schiminovich, D., and

Tumlinson, J. (2017) The LUVOIR Ultraviolet Multi-Object Spectrograph (LUMOS): instrument definition and design. *Proceedings of the SPIE* 10397

Franz, H. B., Trainer, M. G., Malespin, C. A., Mahaffy, P. R., Atreya, S. K., Becker, R. H., Benna, M., Conrad, P. G., Eigenbrode, J. L., Freissinet, C. *et al.* (2017) Initial SAM calibration gas experiments on Mars: Quadrupole mass spectrometer results and implications. *Planet Space Sci* 138:44-54

Fujii, Y., Angerhausen, D., Deitrick, R., Domagal-Goldman, S., Grenfell, J. L., Hori, Y., Kane, S. R., Palle, E., Rauer, H., Siegler, N., Stapelfeldt, K., and Stevenson, K. B. (2018) Exoplanet Biosignatures: Observational Prospects. *Astrobiology* 18

Galloway, J. N., Schlesinger, W. H., Levy, H. I., Michaels, A., and Schnoor, J. L. (1995) Nitrogen fixation: Anthropogenic enhancement-environmental response. *Global Biochemical Circles* 9:235-252

Galloway, J. N. (2003) The Global Nitrogen Cycle. In *Treatise on geochemistry* 8, H. Holland and K. Turekian, Elsevier Limited, Amsterdam, pp 557-583

Galloway, J. N. (2003) The Global Nitrogen Cycle. In *Treatise on geochemistry* 8, H. Holland and K. Turekian, Elsevier Limited, Amsterdam, pp 557-583

Gao, P., Hu, R., Robinson, T. D., Li, C., and Yung, Y. L. (2015) Stability of $CO_2$ Atmospheres on Desiccated M Dwarf Exoplanets. *Astrophys J* 806

García Muñoz, A., Zapatero Osorio, M. R., Barrena, R., Montañés-Rodríguez, P., Martín, E. L., and Pallé, E. (2012) Glancing Views of the Earth: From a Lunar Eclipse to an Exoplanetary Transit. *Astron J* 755

Gershberg, R. E., Katsova, M. M., Lovkaya, M. N., Terebizh, A. V., and Shakhovskaya, N. I. (1999) Catalogue and bibliography of the UV Cet-type flare stars and related objects in the solar vicinity. *Astron Astrophys Suppl* 139:555-558

Gerya, T. V., Stern, R. J., Baes, M., Sobolev, S. V., and Whattam, S. A. (2015) Plate tectonics on the Earth triggered by plume-induced subduction initiation.. *Nature* 527:221-225

Gillmann, C., Chassefière, E., and Lognonné, P. (2009) A consistent picture of early hydrodynamic escape of Venus atmosphere explaining present Ne and Ar isotopic ratios and low oxygen atmospheric content. *Earth Planet Sci Lett* 286:503-513

Goldblatt, C., Claire, M. W., Lenton, T. M., Matthews, A. J., Watson, A. J., and Zahnle, K. J. (2009) Nitrogen-enhanced greenhouse warming on early Earth. *Nat Geosci* 2:891-896

Grenfell, J. L., Rauer, H., Selsis, F., Kaltenegger, L., Beichman, C., Danchi, W., Eiroa, C., Fridlund, M., Henning, T., Herbst, T. *et al.* (2010) Co-Evolution of Atmospheres, Life, and Climate. *Astrobiology* 10:77-88

Grenfell, J. L., Grießmeier, J.-M., Patzer, B., Rauer, H., Segura, A., Stadelmann, A., Stracke, B., Titz, R., and Von Paris, P. (2007a) Biomarker Response to Galactic Cosmic Ray-Induced NOx And The Methane Greenhouse Effect in The Atmosphere of An Earth-Like Planet Orbiting An M Dwarf Star. *Astrobiology* 7:208-221

Grenfell, J. L., Stracke, B., von Paris, P., Patzer, B., Titz, R., Segura, A., and Rauer, H. (2007b) The response of atmospheric chemistry on earthlike planets around F, G and K Stars to small variations in orbital distance. *Planet Space Sci* 55:661-671

Grenfell, J. L., Grießmeier, J.-M., von Paris, P., Patzer, A. B., Lammer, H., Stracke, B., Gebauer, S., Schreier, F., and Rauer, H. (2012) Response of Atmospheric Biomarkers to NOx-Induced Photochemistry Generated by Stellar Cosmic Rays for Earth-like Planets in the Habitable Zone of M Dwarf Stars. *Astrobiology* 12:1109-1122

Grenfell, J. L., Gebauer, S., Godolt, M., Stracke, B., Lehmann, R., and Rauer, H. (2018) Limitation of atmospheric composition by combustion-explosion in exoplanetary atmospheres. *Astrophys J (accepted)*

Grießmeier, J.-M., Stadelmann, A., Motschmann, U., Belisheva, N. K., Lammer, H., and Biernat, H. K. (2005) Cosmic Ray Impact on Extrasolar Earth-Like Planets in Close-in Habitable Zones. *Astrobiology* 5:587-603

Grießmeier, J.-M., Tabataba-Vakili, F., Stadelmann, A., Grenfell, J. L., and Atri, D. (2016) Galactic cosmic rays on extrasolar Earth-like planets. II. Atmospheric implications. *Astron Astrophys* 587

Hacker, B. R., B., K. P., and Behn, M. D. (2011) Differentiation of the continental crust by relamination. *Earth Planet Sci Lett* 307:501-516

Halliday, A. N. (2003) The Origin and Earliest History of the Earth. *Treatise on Geochemistry* 1:509-557

Hamano, K., Abe, Y., and Genda, H. (2013) Emergence of two types of terrestrial planet on solidification of magma ocean. *Nature* 497:607-610

Haqq-Misra, J. D., Domagal-Goldman, S. D., Kasting, P. J., and Kasting, J. F. (2008) A Revised, Hazy Methane Greenhouse for the Archean Earth. *Astrobiology* 8:1127-1137

Harman, C. E., Schwieterman, E. W., Schottelkotte, J. C., and Kasting, J. F. (2015) Abiotic O2 Levels on Planets around F, G, K, and M Stars: Possible False Positives for Life?. *Astrophys J* 812

Hedelt, P., von Paris, P., Godolt, M., Gebauer, S., Grenfell, J. L., Rauer, H., Schreier, F., Selsis, F., and Trautmann, T. (2013) Spectral features of Earth-like planets and their detectability at different orbital distances around F, G, and K-type stars. *Astron Astrophys* 553

Hessler, A. M., Lowe, D. R., Jones, R. L., and Bird, D. K. (2004) A lower limit for atmospheric carbon dioxide levels 3.2 billion years ago. *Nature* 428:736-738

Hilton, D. R., Fischer, T. P., and Marty, B. (2002) Noble Gases and Volatile Recycling at Subduction Zones. In *Noble Gases in Geochemistry and Cosmochemistry* 47 (Reviews in mineralogy and geochemistry), D. Porcelli, C. J. Ballentine, and R. Wieler, Mineralogical Society of America, pp 319-370

Hirschmann, M. M. (2009) Partial melt in the oceanic low velocity zone. *Phys Earth Planet Inter* 179:60-71

Holland, H. D. (1962) Petrologic studies. In *A volume to honor* A. F. Buddington, E. J. Engel, H. L. James, and B. F. Leonard, Geological Society of America, New York, pp 447-477

Holland, H. D. (1978), The chemistry of the atmosphere and oceans Wiley, New York, p 351

Holland, H. D. (1984), The chemical evolution of the atmosphere and oceans Princeton Univ Press, Princeton

Holloway, J. M. and Dahlgren, R. A. (2002) Nitrogen in rock: Occurrences and biogeochemical implications. *Global Biogeochem Cycles* 16:65:1-65:17

Höning, D., Hansen-Goos, H., Airo, A., and Spohn, T. (2014) Biotic vs. abiotic Earth: A model for mantle hydration and continental coverage. *Planet Space Sci* 98:5-13

Höning, D. and Spohn, T. (2016) Continental growth and mantle hydration as intertwined feedback cycles in the thermal evolution of Earth. *Phys Earth Planet Inter* 255:27-49

Hopkins, M. D., Harrison, M., and Manning, C. E. (2010) Constraints on Hadean geodynamics from mineral inclusions in > 4 Ga zircons. *Earth Planet Sci Lett* 298:367-376

Hopkins, M. T. and Manning, C. E. (2008) Low heat flow inferred from >4 Gyr zircons suggests Hadean plate boundary interactions. *Nature* 456:493-496

Jacob, D. J. (1999), Atmospheric Chemistry Princeton University Press

Jaffe, D. A. (1992) The Nitrogen Cycle. In *Global Biochemical Cycles* S. S. Butcher, R. J. Charlson, G. H. Orians, and G. V. Wolfe, Academic Press Limited, London, pp 263-284

Johnson, B. W. and Goldblatt, C. (2018) EarthN: A new Earth System Nitrogen Model: submitted. *Geochem Geophys Geosys; arXiv:1805.00893v1*

Johnstone, C. P., Güdel, M., Stökl, A., Lammer, H., Tu, L., Kislyakova, K. G., Lüftinger, T., Odert, P., Erkaev, N. V., and Dorfi, E. A. (2015) The Evolution of Stellar Rotation and the Hydrogen Atmospheres of Habitable-zone Terrestrial Planets. *Astrophys J Lett* 815

Jones, H. (2003) Searching for Alien Life Having Unearthly Biochemistry. *SAE Technical Paper 2003-01-2668*

Joshi, M. (2003) Climate Model Studies of Synchronously Rotating Planets. *Astrobiology* 3:415-427

Kadik, A. A., Kurovskaya, N. A., Ignat'ev, Y. A., Kononkova, N. N., Koltashev, V. V., and Plotnichenko, V. G. (2011) Influence of oxygen fugacity on the solubility of nitrogen, carbon, and hydrogen in FeO-Na2O-SiO2-Al2O3 melts in equilibrium with metallic iron at 1.5 GPa and 1400°C. *Geochem Int* 49

Kaltenegger, L., Traub, W. A., and Jucks, K. W. (2007) Spectral Evolution of an Earth-Like Planet. *Astrophys J* 658:598-616

Kanzaki, Y. and Murakami, T. (2015) Estimates of atmospheric CO2 in the Neoarchean-Paleoproterozoic from paleosols. *Geochimica et Cosmochimica Acta* 159:190-219

Kasting, J. F. (1982) Stability of ammonia in the primitive terrestrial atmosphere. *J Geophys Res* 87:3091-3098

Kasting, J. F. (1993) Earth's early atmosphere. *Nature* 259:920-926

Kasting, J. F., Whitmire, D. P., and Reynolds, R. T. (1993) Habitable Zones around Main Sequence Stars. *Icarus* 101:108-128

Kelley, K. A. and Cottrell, E. (2009) Water and the Oxidation State of Subduction Zone Magmas. *Science* 325:605-607

Kharecha, P., Kasting, J., and J. Siefert (2005) A coupled atmosphere-ecosystem model of the early Archean Earth. *Geobiology* 3:53-76

Khodachenko, M. L., Ribas, I., Lammer, H., Grießmeier, J.-M., Leitner, M., Selsis, F., Eiroa, C., Hanslmeier, A., Biernat, H. K., Farrugia, C. J., and Rucker, H. O. (2007) Coronal Mass Ejection (CME) Activity of Low Mass M Stars as An Important Factor for The Habitability of Terrestrial Exoplanets. I. CME Impact on Expected Magnetospheres of Earth-Like Exoplanets in Close-In Habitable Zones. *Astrobiology* 7:167-184

Kiehl, J. T. and Dickinson, R. E. (1987) A study of the radiative effects of enhanced atmospheric $CO_2$ and $CH_4$ on early Earth surface temperatures. *J Geophys Res* 92:2991-2998

Kislyakova, K. G., Noack, L., Johnstone, C. P., Zaitsev, V. V., Fossati, L., Lammer, H., Khodachenko, M. L., Odert, P., and Güdel, M. (2017) Magma oceans and enhanced volcanism on TRAPPIST-1 planets due to induction heating. *Nat Astron* 1:878-885

Kislyakova, K. G., Fossati, L., Johnstone, C. P., Noack, L., Lueftinger, T., Zaitsev, V. V., and Lammer, H. (2018) Effective induction heating around strongly magnetized stars. *Astrophys J* 858

Kitzmann, D., Alibert, Y., Godolt, M., Grenfell, J. L., Heng, K., Patzer, B., Rauer, H., Stracke, B., and von Paris, P. (2015) The unstable CO2 feedback cycle on ocean planets. *Mon Not R Astron* 452:3752-3758

Korenaga, J. (2013) Initiation and Evolution of Plate Tectonics on Earth: Theories and Observations. *Annu Rev Earth Planet Sci* 41:117-151

Krasnopolsky, V. A., Maillard, J. P., and Owen, T. C. (2004) Detection of methane in the martian atmosphere: evidence for life?. *Icarus* 172:537-547

Krissansen-Totton, J., Bergsman, D. S., and Catling, D. C. (2016a) On Detecting Biospheres from Chemical Thermodynamic Disequilibrium in Planetary Atmospheres. *Astrobiology* 16:39-67

Krissansen-Totton, J., Schwieterman, E. W., Charnay, B., Arney, G., Robinson, T. D., Meadows, V., and Catling, D. C. (2016b) Is the Pale Blue Dot Unique? Optimized Photometric Bands for Identifying Earth-like Exoplanets. *Astrophys J* 817:1-20

Krissansen-Totton, J. and Catling, D. C. (2017) Constraining climate sensitivity and continental versus seafloor weathering using an inverse geological carbon cycle model. *Nat Commun* 8

Krissansen-Totton, J., Arney, G. N., and Catling, D. C. (2018a) Constraining the climate and ocean pH of the early Earth with a geological carbon cycle model. *Proc Natl Acad Sci USA*

Krissansen-Totton, J., Olson, S., and Catling, D. C. (2018b) Disequilibrium biosignatures over Earth history and implications for detecting exoplanet life. *Sci Adv* 4

Kuhn, W. R. and Atreya, S. K. (1979) Ammonia photolysis and the greenhouse effect in the primordial atmosphere of the earth. *Icarus* 37:207-213

Kump, L. R., Kasting, J. F., and Barley, M. E. (2001) Rise of atmospheric oxygen and the "upside-down" Archean mantle. *Geochem Geophys Geosyst* 2:1025-1027

Kunze, M., Godolt, M., Langematz, U., Grenfell, J. L., Hamann-Reinus, A., and Rauer, H. (2014) Investigating the early Earth faint young Sun problem with a general circulation model. *Planet Space Sci* 98:77-92

Kurosawa, K. (2015) Impact-driven planetary desiccation: The origin of the dry Venus. *Earth Planet Sci Lett* 429:181-190

Lammer, H., Bredehöft, J. H., Coustenis, A., Khodachenko, M. L., Kaltenegger, L., Grasset, O., Prieur, D., Raulin, F., Ehrenfreund, P., Yamauchi, M. *et al.* (2009) What makes a planet habitable?. *Astron Astrophys Review* 17:181-249

Lammer, H., Stökl, A., Erkaev, N. V., Dorfi, E. A., Odert, P., Güdel, M., Kulikov, Y. N., Kislyakova, K. G., and Leitzinger, M. (2014) Origin and loss of nebula-captured hydrogen envelopes from `sub'- to `super-Earths' in the habitable zone of Sun-like stars. *Mon Not R Astron* 439:3225-3238

Lammer, H., Lichtenegger, H. I., Kulikov, Y. N., Grießmeier, J.-M., Terada, N., Erkaev, N. V., Biernat, H. K., Khodachenko, M. L., Ribas, I., Penz, T., and Selsis, F. (2007) Coronal Mass Ejection (CME) Activity of Low Mass M Stars as An Important Factor for The Habitability of Terrestrial Exoplanets. II. CME-Induced

Ion Pick Up of Earth-like Exoplanets in Close-In Habitable Zones. *Astrobiology* 7:185-207

Lammer, H., Kasting, J. F., Chassefière, E., Johnson, R. E., Kulikov, Y. N., and Tian, F. (2008) Atmospheric Escape and Evolution of Terrestrial Planets and Satellites. *Space Sci Rev* 139:399-436

Lammer, H., Kislyakova, K. G., Odert, P., Leitzinger, M., Schwarz, R., Pilat-Lohinger, E., Kulikov, Y. N., Khodachenko, M. L., Güdel, M., and Hanslmeier, A. (2011) Pathways to Earth-Like Atmospheres. Extreme Ultraviolet (EUV)-Powered Escape of Hydrogen-Rich Protoatmospheres. *Orig Life Evol Biosph* 41:503-522

Lammer, H., Kislyakova, K. G., Güdel, M., Holmström, M., Erkaev, N. V., Odert, P., and Khodachenko, M. L. (2013) Stability of Earth-Like N2 Atmospheres: Implications for Habitability. In *The Early Evolution of the Atmospheres of Terrestrial Planets. Astrophysics and Space Science Proceedings* 35, T.-R. J., R. F., M. C., and N. C., Springer Science+Business Media, New York, pp 33-52

Lammer, H., Chassefière, E., Karatekin, Ö., Morschhauser, A., Niles, P. B., Mousis, O., Odert, P., Möstl, U. V., Breuer, D., Dehant, V. *et al.* (2013a) Outgassing History and Escape of the Martian Atmosphere and Water Inventory. *Space Sci Rev* 174:113-154

Lammer, H. and Blanc, M. (2018) From Disks to Planets: The Making of Planets and Their Early Atmospheres. An Introduction. *Space Sci Rev* 214:1-35

Lammer, H., Zerkle, A. L., Gebauer, S., Tosi, N., Noack, L., Scherf, M., Pilat-Lohinger, E., Güdel, M., Grenfell, J. L., Godolt, M., and Nikolaou, A. (2018) Origin and evolution of the atmospheres of early Venus, Earth and Mars. *Astron. Astrophys. Rev.* 26:1-72

Laneuville, M., Kameya, M., and Cleaves, H. J. (2018) Earth Without Life: A Systems Model of a Global Abiotic Nitrogen Cycle. *Astrobiology* 18:897-914

Lebrun, T., Massol, H., ChassefièRe, E., Davaille, A., Marcq, E., Sarda, P., Leblanc, F., and Brandeis, G. (2013) Thermal evolution of an early magma ocean in interaction with the atmosphere. *J Geophys Res Planets* 118:1155-1176

Lee, C.-T. A., Yeung, L. Y., McKenzie, N. R., Yokoyama, Y., Ozaki, K., and Lenardic, A. (2016) Two-step rise of atmospheric oxygen linked to the growth of continents. *Nat Geosci* 9:417-424

Léger, A., Pirre, M., and Marceau, F. J. (1993) Search for primitive life on a distant planet: relevance of O2 and O3 detections. *Astron. Astrophys.* 277:309-313

Léger, A., Fontecave, M., Labeyrie, A., Samuel, B., Demangeon, O., and Valencia, D. (2011) Is the Presence of Oxygen on an Exoplanet a Reliable Biosignature?. *Astrobiology* 11:335-341

Lehmer, O. and Catling, D. (2017) Rocky Worlds Limited to ~1.8 Earth Radii by Atmospheric Escape During a Star's Extreme UV Saturation. *Astrophys J* 845:1-7

Li, Y., Wiedenbeck, M., Shcheka, S., and Keppler, H. (2013) Nitrogen solubility in upper mantle minerals. *Earth Planet Sci Lett* 377:311-323

Libourel, G., Marty, B., and Humbert, F. (2003) Nitrogen solubility in basaltic melt. Part I. Effect of oxygen fugacity. *Geochim Cosmochim Acta* 67:4123-4135

Lichtenegger, H. I., Lammer, H., Grießmeier, J.-M., Kulikov, Y. N., von Paris, P., Hausleitner, W., Krauss, S., and Rauer, H. (2010) Aeronomical evidence for higher CO2 levels during Earth's Hadean epoch. *Icarus* 210

Lichtenegger, H. I., Kislyakova, K. G., Odert, P., Erkaev, N. V., Lammer, H., Gröller, H., Johnstone, C. P., Elkins-Tanton, L., Tu, L., Güdel, M., and Holmström, M. (2016) Solar XUV and ENA-driven water loss from early Venus' steam atmosphere. *J Geophys Res Space Phys* 121:4718-4732

Lovelock, J. E. (1975) Thermodynamics and the recognition of alien biospheres. *Proc. R. Soc. Lond.* 189:167-181

Lovelock, J. E. and Margulis, L. (1974) Atmospheric homeostasis by and for the biosphere: the gaia hypothesis. *Tellus* XXVI

Loyd, R. O., France, K., Youngblood, A., Schneider, C., Brown, A., Hu, R., Linsky, J., Froning, C. S., Redfield, S., Rugheimer, S., and Tian, F. (2016) The MUSCLES Treasury Survey. III. X-Ray to Infrared Spectra of 11 M and K Stars Hosting Planets. *Astrophys J* 824

Luger, R. and Barnes, R. (2015) Extreme Water Loss and Abiotic O2 Buildup on Planets Throughout the Habitable Zones of M Dwarfs. *Astrobiology* 15:119-143

Luger, R., Barnes, R., Lopez, E., Forthey, J., Jackson, B., and Meadows, V. (2015) Habitable Evaporated Cores: Transforming Mini-Neptunes into Super-Earths in the Habitable Zones of M Dwarfs. *Astrobiology* 15:57-88

Lyons, T. W., Reinhard, C. T., and Planavsky, N. J. (2014) The rise of oxygen in Earth's early ocean and atmosphere. *Nature* 315:307

Mallik, A., Li, Y., and Wiedenbeck, M. (2018) Nitrogen evolution within the Earth's atmosphere–mantle system assessed by recycling in subduction zones. *Earth and Planetary Science Letters* 482:556-566

Mandt, K. E., Jr., J. H., Lewis, W., Magee, B., Bell, J., Lunine, J., Mousis, O., and Cordier, D. (2009) Isotopic evolution of the major constituents of Titan's atmosphere based on Cassini data. *Planetary and Space Science* 57:1917-1930

Mandt, K. E., Mousis, O., Lunine, J., and Gautier, D. (2014) Protosolar Ammonia as the Unique Source of Titan's nitrogen. *Astrophys J Lett* 788

Margulis, L. and Lovelock, J. E. (1974) Biological Modulation of the Earth's Atmosphere. *Icarus* 21:471-789

Martin, R. S., Mather, T. A., and Pyle, D. M. (2007) Volcanic emissions and the early Earth atmosphere. *Geochim Cosmochim Acta* 71:3673-3685

Marty, B. (1995) Nitrogen content of the mantle inferred from N2-Ar correlation in oceanic basalts. *Nature* 377:326-329

Marty, B. and Tolstikhin, I. (1998) CO2 fluxes from mid-ocean ridges, arcs and plumes. *Chem Geology* 145:233-248

Marty, B., Zimmermann, L., Pujol, M., Burgess, R., and Philippot, P. (2013) Nitrogen Isotopic Composition and Density of the Archean Atmosphere. *Science* 342:101-104

Massol, H., Hamano, K., Tian, F., Ikoma, M., Abe, Y., Chassefière, E., Davaille, A., Genda, H., Güdel, M., Hori, Y. *et al.* (2016) Formation and Evolution of Protoatmospheres. *Space Sci Rev* 205:153-211

Mayor, M. and Queloz, D. (1995) A Jupiter-mass companion to a solar-type star. *Nature* 378:335-359

McCammon, C. (2005) The Paradox of Mantle Redox. *Science* 308:807-808

Meadows, V. S., Reinhard, C. T., Arney, G. N., Parenteau, M. N., Schwieterman, E. W., Domagal-Goldman, S. D., Lincowski, A. P., Stapelfeldt, K. R., Rauer, H., DasSarma, S., Hegde, S., Narita, N., and Deitr (2017) Exoplanet Biosignatures: Understanding Oxygen as a Biosignature in the Context of Its Environment. *submitted to Astrobiology, eprint arXiv:1705.07560*

Mennesson, B., Gaudi, S., Seager, S., Cahoy, K., Domagal-Goldman, S., Feinberg, L., Guyon, O., Kasdin, N., Marois, C., Mawet, D. *et al.* (2016) The Habitable Exoplanet (HabEx) Imaging Mission: preliminary science drivers and technical requirements. *Space Telescopes and Instrumentation 2016*, eds MacEwen, H. A., Fazio, G. G., and Lystrup, M., Proc. of SPIE Vol. 9904

Mikhail, S. and Sverjensky, D. (2014) Nitrogen speciation in upper mantle fluids and the origin of Earth's nitrogen-rich atmosphere. *Nature Geoscience* 7:816-819

Misra, A., Meadows, V., Claire, M., and Crisp, D. (2014a) Using dimers to measure biosignatures and atmospheric pressure for terrestrial exoplanets. *Astrobiology* 14:67-86

Misra, A., Meadows, V., and Crisp, D. (2014b) The effects of refraction on transit transmission spectroscopy: application to earth-like exoplanets. *Astrophys J* 792

Muller, C. (2013) N2O as a Biomarker, from the Earth and Solar System to Exoplanets. In *The Early Evolution of the Atmospheres of Terrestrial Planets. Astrophysics and Space Science Proceedings* 35, T.-R. J., R. F., M. C., and N. C., Springer Science+Business Media, New York, pp 99-106

NASA (2017) *Earth Fact Sheet*. Retrieved June *4, 2018*: http://nssdc.gsfc.nasa.gov/planetary/factsheet/earthfact.html

Navarro-González, R., Molina, M. J., and Molina, L. T. (1998) Nitrogen fixation by volcanic lightning in the early Earth. *Geophys Res Lett* 25:3123-3126

Navarro-González, R., McKay, C. P., and Mvondo, D. N. (2001) A possible nitrogen crisis for Archean life due to reduced nitrogen fixation by lightning. *Nature* 412:61-64

Noack, L., Höning, D., Rivoldini, A., Heistracher, C., Zimov, N., Journaux, B., Lammer, H., Van Hoolst, T., and Bredehöft, J. H. (2016) Water-rich planets: How habitable is a water layer deeper than on Earth?. *Icarus* 277:215-236

O'Neill, C., Lenardic, A., Weller, M., Moresi, L., Quenette, S., and Zhang, S. (2016) A window for plate tectonics in terrestrial planet evolution?. *Phys Earth Planet Inter* 255:80-92

Odert, P., Lammer, H., Erkaev, N. V., Nikolaou, A., Lichtenegger, H. I., Johnstone, C. P., Kislyakova, K. G., Leitzinger, M., and Tosi, N. (2018) Escape and fractionation of volatiles and noble gases from Mars-sized planetary embryos and growing protoplanets. *Icarus* 307:327-346

Owen, J. E. and Mohanty, S. (2016) Habitability of terrestrial-mass planets in the HZ of M Dwarfs - I. H/He-dominated atmospheres. *Mon Not R Astron Soc* 459:4088-4108

Owen, T. (1980) The Search for Early Forms of Life in Other Planetary Systems: Future Possibilities Afforded by Spectroscopic Techniques. In *Strategies for the Search for Life in the Universe. Astrophysics and Space Science Library (A Series of Books on the Recent Developments of Space Science and of General Geophysics and Astrophysics)* 83, M. D. Papagiannis, Springer, Dordrecht, pp 177-185

Oyama, V. I., Carle, G. C., Woeller, F., Pollack, J. B., Reynolds, R. T., and Craig, R. A. (1980) Pioneer Venus gas chromatography of the lower atmosphere of Venus. *J Geophys Res* 85:7891-7902

Pallé, E., Zapatero Osorio, M. R., Barrena, R., Montañés-Rodríguez, P., and Martín, E. L. (2009) Earth's transmission spectrum from lunar eclipse observations. *Nature* 459:814-816

Parkos, D., Pikus, A., Alexeenko, A., and Melosh, H. J. (2016) HCN production from impact ejecta on early Earth. *AIP Conference Proceedings* 1786, 170001

Parkos, D., Pikus, A., Alexeenko, A., and Melosh, H. J. (2018) HCN Production via Impact Ejecta Reentry During the Late Heavy Bombardment. *J Geophys Res* 123:892-909

Pavlov, A. A., Vasilyev, G., Ostryakov, V. M., Pavlov, A. K., and Mahaffy, P. (2012) Degradation of the organic molecules in the shallow subsurface of Mars due to irradiation by cosmic rays. *Geophys Res Lett* 39:1-5

Pearson, D. G., Brenker, F. E., Nestola, F., McNeill, J., Nasdala, L., Hutchison, M. T., Matveev, S., Mather, K., Silversmit, G., Schmitz, S., Vekemans, B., and Vincze, L. (2014) Hydrous mantle transition zone indicated by ringwoodite included within diamond. *Nature* 507:221-224

Pierrehumbert, R. T. (2010), Principles of Planetary Climate Cambridge Univ Press, Cambridge

Pilcher, C. B. (2003) Biosignatures of Early Earths. *Astrobiology* 3:471-486

Plümper, O., King, H. E., Geisler, T., Liu, Y., Pabst, S., Savov, I. P., Rost, D., and Zack, T. (2017) Subduction zone forearc serpentinites as incubators for deep microbial life. *Proc Natl Acad Sci* 114:4324-4329

Rakov, V. A. and Uman, M. A. (2003), Lightning Cambridge Univ Press, Cambridge

Rakov, V. A. and Uman, M. A. (2004), Lightning: Physics and effects Cambridge University Press, Cambridge

Rauer, H., Gebauer, S., Paris, P. V., Cabrera, J., Godolt, M., Grenfell, J. L., Belu, A., Selsis, F., Hedelt, P., and Schreier, F. (2011) Potential biosignatures in super-Earth atmospheres. I. Spectral appearance of super-Earths around M dwarfs. *Aston Astrophys* 529

Rauer, H., Catala, C., Aerts, C., Appourchaux, T., Benz, W., Brandeker, A., Christensen-Dalsgaard, J., Deleuil, M., Gizon, L., Goupil, M.-J. *et al.* (2014) The PLATO 2.0 mission. *Exp Astron* 38:249-330

Ronov, A. B. and Yaroshevskiy, A. A. (1967) Chemical strucutre of the Earth's crust. *Geochemistry* 11:1041-1066

Rosswall, T. (1983) The nitrogen cycle. In *The major Biogeochemical Cycles and Their Interactions* B. Bolin and R. B. Cook, John Wiley & Sons, Chichester, pp 46-50

Rugheimer, S., Kaltenegger, L., Segura, A., Linsky, J., and Mohanty, S. (2015) Effect of UV Radiation on the Spectral Fingerprints of Earth-like Planets Orbiting M Stars. *Astrophys J* 809

Rugheimer, S., Kaltenegger, L., Zsom, A., Segura, A., and Sasselov, D. (2013) Spectral Fingerprints of Earth-like Planets Around FGK Stars. *Astrobiology* 13:251-269

Sagan, C., Thompson, W. R., Carlson, R., Gurnett, D., and Hord, C. (1993) A search for life on Earth from the Galileo spacecraft. *Nature* 365:715-721

Salvador, A., Massol, H., Davaille, A., Marcq, E., Sarda, P., and Chassefière, E. (2017) The relative influence of $H_2O$ and $CO_2$ on the primitive surface conditions and evolution of rocky planets. *J Geophys Res Planet* 122:1458-1486

Sano, Y., Takahata, N., Nishio, Y., Fischer, T. P., and Williams, S. N. (2001) Volcanic flux of nitrogen from the Earth. *Chemical Geology* 171:263-271

Sasaki, S. and Nakazawa, K. (1988) Origin of isotopic fractionation of terrestrial Xe: hydrodynamic fractionation during escape of the primordial H 2sbnd He atmosphere. *Earth Planet Sci Lett* 89:323-334

Scalo, J., Kaltenegger, L., Segura, A. G., Fridlund, M., Ribas, I., Kulikov, Y. N., Grenfell, J. L., Rauer, H., Odert, P., Leitzinger, M. *et al.* (2007) M Stars as Targets for Terrestrial Exoplanet Searches And Biosignature Detection. *Astrobiology* 7:85-166

Schaefer, L. and Fegley, B. J. (2010) Chemistry of atmospheres formed during accretion of the Earth and other terrestrial planets. *Icarus* 208:438-448

Scherf, M., Khodachenko, M., Blokhina, M., Johnstone, C., Alexeev, I., Belenkaya, E., Tarduno, J., Tu, L., Lichtenegger, H., Guedel, M., and Lammer, H. (2018) On the Earth's paleo-magnetosphere during the late Hadean eon and possible implications for the ancient terrestrial atmosphere. *Earth Planet Sci Lett: submitted*

Schirrmeister, B. E., Gugger, M., and Donoghue, P. C. (2015) Cyanobacteria and the Great Oxidation Event: Evidence from Genesis and Fossils. *Palaeontology* 58:769-785

Schlesinger, W. H. (1997), Biochemistry: An Analysis of Global Change Elsevier, New York

Schmandt, B., Jacobsen, S. D., Becker, T. W., Liu, Z., and Dueker, K. G. (2014) Dehydration melting at the top of the lower mantle. *Sciencce* 344:1265-1268

Schopf, J. W. (2014) Geological evidence of oxygenic photosynthesis and the biotic response to the 2400-2200 Ma "Great Oxidation Event". *Biochemistry* 79:165-177

Schwieterman, E. W., Kiang, N. Y., Parenteau, M. N., Harman, C. E., DasSarma, S., Fisher, T. M., Arney, G. N., Hartnett, H. E., Reinhard, C. T., Olson, S. L. *et al.* (2018) Exoplanet Biosignatures: A Review of Remotely Detectable Signs of Life. *Astrobiology* 18

Seager, S., Whitney, B. A., and Sasselov, D. D. (2000) Photometric Light Curves and Polarization of Close-in Extrasolar Giant Planets. *Astrophys J* 540:504-520

Seager, S., Bains, W., and Hu, R. (2013a) A Biomass-based Model to Estimate the Plausibility of Exoplanet Biosignature Gases. *Astrophys J* 775

Seager, S., Bains, W., and Hu, R. (2013b) Biosignature Gases in H2-dominated Atmospheres on Rocky Exoplanets. *Astrophys J* 777

Segura, A., Kasting, J. F., Meadows, V., Cohen, M., Scalo, J., Crisp, D., Butler, R. A., and Tinetti, G. (2005) Biosignatures from Earth-Like Planets Around M Dwarfs. *Astrobiology* 5:706-725

Segura, A., Krelove, K., Kasting, J. F., Sommerlatt, D., Meadows, V., Crisp, D., Cohen, M., and Mlawer, E. (2003) Ozone Concentrations and Ultraviolet Fluxes on Earth-Like Planets Around Other Stars. *Atrobiology* 3:689-708

Sekiya, M., Nakazawa, K., and Hayashi, C. (1980a) Dissipation of the Primordial Terrestrial Atmosphere Due to Irradiation of the Solar EUV. *Prog Theoret Phys* 64:1968-1985

Sekiya, M., Nakazawa, K., and Hayashi, C. (1980b) Dissipation of the rare gases contained in the primordial earth's atmosphere. *Earth Planet Sci Lett* 50:197-201

Shields, A. L., Meadows, V. S., Bitz, C. M., Pierrehumbert, R. T., Joshi, M. M., and Robinson, T. D. (2013) The Effect of Host Star Spectral Energy Distribution and Ice-Albedo Feedback on the Climate of Extrasolar Planets. *Astrobiology* 13:715-739

Shirey, S. B., Kamber, B. S., Whitehouse, M. J., Mueller, P. A., and Basu, A. R. (2008) A review of the isotopic and trace element evidence for mantle and crustal processes in the Hadean and Archean: Implications for the onset of plate tectonic subduction. *Geol Soc Spec* 440

Sleep, N. H. and Zahnle, K. (2001) Carbon dioxide cycling and implications for climate on ancient Earth. *J Geophys Res Atmos* 106:1373-1399

Sleep, N. H. (2010) The Hadean-Archaean Environment. *Cold Spring Harb Perspect Biol* 2: a002527

Smith, D. S. and Scalo, J. W. (2004) Transprot of ionishing radiation in terrestrial-like exoplanet atmospheres. *Icarus* 171:229-253

Snellen, I. A., Désert, J.-M., Waters, L. B., Robinson, T., Meadows, V., van Dishoeck, E. F., Brandl, B. R., Henning, T., Bouwman, J., Lahuis, F. *et al.* (2017) Detecting Proxima b's Atmosphere with JWST Targeting CO2 at 15 μm Using a High-pass Spectral Filtering Technique. *Astron J* 154

Solomatov, V. S. (2000) Fluid dynamics of a terrestrial magma ocean. In *Origin of the Earth and Moon* 1, R. Canup and K. Righter, University of Arizona Press, Tucson, AZ, pp 323-338

Som, S. M., Catling, D. C., Harnmeijer, J. P., Polivka, P. M., and Buick, R. (2012) Air density 2.7 billion years ago limited to less than twice modern levels by fossil raindrop imprints. *Nature* 484:359-362

Som, S. M., Buick, R., Hagadorn, J. W., Blake, T. S., Perreault, J. M., Harnmeijer, J. P., and Catling, D. C. (2016) Earth's air pressure 2.7 billion years ago constrained to less than half of modern levels. *Nat Geosci* 9:448-451

Southam, G., Westall, F., and Spohn, T. (2015) Geology, life and habitability. In *Treatise on Geophysics* 2nd edition, pp 473-486

Stedman, D. H. and Shetter, R. E. (1983) The Global Budget of Atmospheric Nitrogen Species. In *Trace Atmospheric Constituents* S. E. Schwartz, John Wiley & Sons, New York, pp 411-454

Strobel, D. F. and Shemansky, D. E. (1982) EUV emission from Titan's upper atmosphere: Voyager 1 encounter. *J Geophys Res* 87:1361-1368

Stüeken, E. E., Kipp, M. A., Koehler, M. C., Schwieterman, E. W., Johnson, B., and Buick, R. (2016a) Modeling pN2 through Geological Time: Implications for Planetary Climates and Atmospheric Biosignatures. *Astrobiology* 16:949-963

Stüeken, E. E., Kipp, M. A., Koehler, M. C., and Buick, R. (2016b) The evolution of Earth's biogeochemical nitrogen cycle. *Earth Sci Rev* 160:220-239

Tabataba-Vakili, F., Grenfell, J. L., Grießmeier, J.-M., and Rauer, H. (2016) Atmospheric effects of stellar cosmic rays on Earth-like exoplanets orbiting. *Astron Astrophys* 585

Tackley, P. J. (2000) Mantle convection and plate tectonics: toward an integrated physical and chemical theory. *Science* 288:2002-2007

Tian, F., Kasting, J. F., and Zahnle, K. (2011) Revisiting HCN formation in Earth's early atmosphere. *Earth Planet Sci Lett* 308:417-423

Tian, F., Kasting, J. F., Liu, H.-L., and Roble, R. G. (2008a) Hydrodynamic planetary thermosphere model: 1. Response of the Earth's thermosphere to extreme solar EUV conditions and the significance of adiabatic cooling. *Journal of Geophysical Research* 113

Tian, F., Solomon, S. C., Qian, L., Lei, J., and Roble, R. G. (2008b) Hydrodynamic planetary thermosphere model: 2. Coupling of an electron transport/energy deposition model. *Journal of Geophysical Research* 113

Tian, F., France, K., Linsky, J. L., Mauas, P. J., and Vieytes, M. C. (2014) High stellar FUV/NUV ratio and oxygen contents in the atmospheres of potentially habitable planets. *Earth Planet Sci Lett* 385:22-27

Tian, W. (2009) Thermal escape from super earth atmospheres in the habitable zones of M stars. *Astrophys J* 703:905-909

Tosi, N., Godolt, M., Stracke, B., Ruedas, T., Grenfell, L., Höning, D., Nikolaou, A., Plesa, A. C., Breuer, D., and Spohn, T. (2017) The Habitability of a Stagnant-Lid Earth. *Astron Astrophys* 605

Trail, D., B., W. E., and Tailby, N. D. (2011) The oxidation state of Hadean magmas and implications for early Earth's atmosphere. *Nature* 480:79-82

Tu, L., Johnstone, C. P., Güdel, M., and Lammer, H. (2015) The extreme ultraviolet and X-ray Sun in Time: High-energy evolutionary tracks of a solar-like star. *Astron Astrophys* 577

Turner, G., Burgess, R., and Bannon, M. (1990) Volatile-rich mantle fluids inferred from inclusions in diamond and mantle xenoliths. *Nature* 344:653-655

van Berk, W., Fu, Y., and Ilger, J.-M. (2012) Reproducing early Martian atmospheric carbon dioxide partial pressure by modeling the formation of Mg-Fe-Ca carbonate identified in the Comanche rock outcrops on Mars. *J Geophys Res* 117

van Hunen, J. and Moyen, J.-F. (2012) Archean Subduction: Fact or Fiction?. *Annu Rev Earth Planet Sci* 40:195-219

Voss, M., Bange, H. W., Dippner, J. W., Middelburg, J. J., Montoya, J. P., and Ward, B. (2013) The marine nitrogen cycle: recent discoveries, uncertainties and the potential relevance of climate change. *Philosophical Transactions Royal Socienty B 368*

Walker, J. C. (1985) Carbon dioxide on the early Earth. *Origins Life Evol Biosphere* 16:117-127

Walker, J. C., Hays, P. B., and Kasting, J. F. (1981) A negative feedback mechanism for the long-term stabilization of Earth's surface temperature. *J Geophys Res* 86:9776-9782

Watenphul, A., Wunder, B., Wirth, R., and Heinrich, W. (2010) Ammonium-bearing clinopyroxene: A potential nitrogen reservoir in the Earth's mantle. *Chem Geol* 270:240-248

Wolf, E. T. and Toon, O. B. (2013) Hospitable Archean Climates Simulated by a General Circulation Model. *Astrobiology* 13:656-673

Wordsworth, R. D. (2016) Atmospheric nitrogen evolution on Earth and Venus. *Earth and Planetary Science Letters* 447:103-111

Wordsworth, R. and Pierrehumbert, R. (2014) Abiotic Oxygen-dominated Atmospheres on Terrestrial Habitable Zone Planets. *Astrophys J* 785

Yang, J., Cowan, N. B., and Abbot, D. S. (2013) Stabilizing cloud feedback dramatically expands the habitable zone of tidally locked planets. *Astrophys J Lett* 771

Youngblood, A., France, K., Loyd, R. O., Linsky, J. L., Redfield, S., Schneider, P. C., Wood, B. E., Brown, A., Froning, C., Miguel, Y., Rugheimer, S., and Walkowicz, L. (2016) The MUSCLES Treasury Survey. II. Intrinsic LYα and Extreme Ultraviolet Spectra of K and M Dwarfs with Exoplanets. *Astrophys J* 824

Zahnle, K. J. (1986) Photochemistry of methane and the formation of hydrocyanic acid (HCN) in the Earth's early atmosphere. *J Geophys Res* 91:2819-2834

Zahnle, K. J. and Kasting, J. F. (1986) Mass fractionation during transonic escape and implications for loss of water from Mars and Venus. *Icarus* 68:462-480

Zahnle, K. J., Catling, D. C., and Claire, M. W. (2013) The rise of oxygen and the hydrogen hourglass. *Chem Geol* 362:26-34

Zerkle, A. L. and Mikhail, S. (2017) The geobiological nitrogen cycle: From microbes to the mantle. *Geobiology* 15:343–352

Zerkle, A. L., Poulton, S. W., Newton, R. J., Mettam, C., Claire, M. W., Bekker, A., and Junium, C. K. (2017) Onset of the aerobic nitrogen cycle during the Great Oxidation Event. *Nature* 542:465-467

Zhang, Y. and Zindler, A. (1993) Distribution and evolution of carbon and nitrogen in Earth. *Earth Planet Sci Lett* 117:331-345

## ACKNOWLEDGEMENTS

H. Lammer, L. Sproß, and M. Scherf acknowledge support by the Austrian Science Fund (FWF) NFN project S11601-N16, "Pathways to Habitability: From Disks to Active Stars, Planets and Life" and the related FWF NFN subprojects S11607-N16 "Particle/Radiative Interactions with Upper Atmospheres of Planetary Bodies under Extreme Stellar Conditions", and S11606-N16 "Magnetospheres". L. Grenfell, acknowledge the collaboration within the COST Action TD 1308.

Finally, the authors thank the anonymous referees for their fruitful comments and suggestions which helped to improve the manuscript.

## AUTHOR DISCLOSURE STATEMENT

No competing financial interests exist.